\documentclass[useAMS,usenatbib]{mn2e}
\usepackage{epsfig}

\title[Echelle Spectrophotometry of S~311]{Deep echelle spectrophotometry
of S~311, a Galactic H~{\sc ii} region located outside the solar circle \thanks{Based on observations collected at the European 
Southern
Observatory, Chile, proposal number ESO 68.C-0149(A)}}
\author[J. Garc\'\i a-Rojas et al.]
  {J.~Garc\'\i a-Rojas,$^1$\thanks{E-mail: jogarcia@iac.es}
  C.~Esteban,$^1$ A.~Peimbert,$^2$ M.~Peimbert,$^2$ 
  \newauthor 
  M.~Rodr\'\i guez,$^3$ M.T.~Ruiz$^4$ \\
  $^1$Instituto de Astrof\'\i sica de Canarias, E-38200 La Laguna, Tenerife, Spain \\
  $^2$Instituto de Astronom\'\i a, UNAM, Apdo. Postal 70-264, M\'exico 04510 D.F., Mexico\\
  $^3$Instituto Nacional de Astrof\'\i sica, \'Optica y Electr\'onica INAOE, Apdo. Postal 51 y 
216, 
  7200 Puebla, Pue., Mexico\\
  $^4$Departamento de Astronom\'\i a, Universidad de Chile, Casilla Postal 36D, Santiago de Chile,
  Chile\\}

\newcommand{\elecd}{$n_{\rm e}$}
\newcommand{\te}{$T_{\rm e}$}
\newcommand{\hb}{H$\beta$}
\newcommand{\ha}{H$\alpha$}

\newcommand{\foiii}{[O~{\sc iii}]}

\newcommand{\foii}{[O~{\sc ii}]}
\newcommand{\fsii}{[S~{\sc ii}]}
\newcommand{\fsiii}{[S~{\sc iii}]}
\newcommand{\fni}{[N~{\sc i}]}
\newcommand{\fnii}{[N~{\sc ii}]}

\newcommand{\fcliii}{[Cl~{\sc iii}]}

\newcommand{\ffeii}{[Fe~{\sc ii}]}
\newcommand{\ffeiii}{[Fe~{\sc iii}]}

\newcommand{\oiii}{O~{\sc iii}}
\newcommand{\nitroi}{N~{\sc i}}
\newcommand{\nii}{N~{\sc ii}}

\newcommand{\sili}{Si~{\sc i}}
\newcommand{\silii}{Si~{\sc ii}}
\newcommand{\oi}{O~{\sc i}}
\newcommand{\oii}{O~{\sc ii}}
\newcommand{\ci}{C~{\sc i}}
\newcommand{\cii}{C~{\sc ii}}

\newcommand{\sii}{S~{\sc ii}}
\newcommand{\siii}{S~{\sc iii}}

\newcommand{\cliii}{Cl~{\sc iii}}
\newcommand{\fariii}{[Ar~{\sc iii}]}
\newcommand{\fei}{Fe~{\sc i}}

\newcommand{\feiii}{Fe~{\sc iii}}

\newcommand{\ari}{Ar~{\sc i}}
\newcommand{\ariii}{Ar~{\sc iii}}

\newcommand{\di}{D\,{\sc i}}
\newcommand{\hi}{H\,{\sc i}}
\newcommand{\hii}{H~{\sc ii}}
\newcommand{\hei}{He~{\sc i}}
\newcommand{\heii}{He~{\sc ii}}

\newcommand{\ts}{\emph{$t^2$}}
\begin{document}

\maketitle

\begin{abstract}

We present echelle spectrophotometry of the Galactic {\hii} region S~311. The data 
have been taken 
with the Very Large Telescope UVES echelle spectrograph in the 3100 to 10400 \AA\ range.
We have measured the intensities of 263 emission lines, 178 are permitted lines of H$^0$, D$^0$ 
(deuterium), 
He$^0$, C$^{0}$, C$^{+}$, N$^{0}$, N$^{+}$, O$^{0}$, O$^{+}$, S$^{+}$, Si$^{0}$, 
Si$^{+}$, Ar$^{0}$ and Fe$^{0}$; some of them are produced by recombination and others mainly by 
fluorescence. Physical conditions have been derived using different continuum and line intensity 
ratios.
We have derived He$^{+}$, C$^{++}$ and O$^{++}$ ionic abundances from pure 
recombination lines as well as abundances from collisionally excited lines for 
a large number of ions of different elements. We have obtained consistent 
estimations of {\ts} applying different methods. We have found that the temperature fluctuations 
paradigm is consistent with the {\te}({\hei}) {\it vs.} {\te}({\hi}) relation for {\hii} 
regions, in contrast with what has been found for planetary nebulae.
We report the detection of deuterium Balmer lines up to D$\delta$ in the blue wings of 
the hydrogen lines, which excitation mechanism seems to be continuum fluorescence.

\end{abstract}


\section{Introduction}\label{intro}

S~311 ---also known as NGC 2467--- is an {\hii} region of the Sharpless catalogue \citep{sharpless59} 
which is located outside the solar circle.
This nebula forms part of the Puppis I association located at 4.0 kpc from the Sun 
\citep{russeil03} and at a
Galactocentric distance of 10.43 kpc (assuming a Galactocentric solar distance of 8.0 kpc). 
\citet{albertetal86} 
concluded that the radio morphology of S~311 is consistent with blister processes.
There are few spectrophotometric studies of S~311 in the literature, most of them forming 
part of analisys involving several {\hii} regions \citep[e.g.][]{hawley78, peimbertetal78, 
kennicuttetal00}. All these works are 
based on the analysis of collisionally excited lines (hereinafter CELs). 

We have taken long-exposure 
high-spectral-resolution spectra with the Very Large Telescope (VLT) UVES echelle 
spectrograph to obtain accurate measurements of very faint permitted lines of heavy element ions 
in S~311. 
We have determined the physical conditions and the chemical abundances of S~311 with high accuracy. 
An important improvement in this work is the 
derivation of C$^{++}$ and O$^{++}$ abundances from several pure recombination lines (hereinafter 
RLs) of {\cii} and {\oii}, making use of high spectral resolution and avoiding the 
problems of line blending.

Traditionally, the abundance studies for {\hii} regions have been based on determinations from 
CELs, whose emissivities are strongly dependent on the temperature 
variations over the observed volume of the nebula. Alternatively, the emissivities of RLs are almost independent of 
such variations and are, in principle, more precise indicators of the true chemical abundances of 
the nebula. 
The use of high resolution echelle spectrographs has permitted our group to obtain deep 
high resolution spectra of bright Galactic {\hii} regions \citep[e.g.][]{estebanetal98, 
estebanetal99a, estebanetal99b, estebanetal04, garciarojasetal04}, and extragalactic 
{\hii} regions \citep[e.g.][]{estebanetal02, apeimbert03}; 
all these works have found that abundance determinations from RLs are systematically larger than 
those obtained 
using CELs (the so called \emph{abundance discrepancy} problem). 
One of the most probable causes of this abundance discrepancy is the presence of spatial 
variations or fluctuations 
in the temperature structure of the nebulae \citep{torrespeimbertetal80}. 
Both phenomena can be related due to the different functional 
dependence of the line emissivities of CELs and RLs on the electron temperature, 
which is stronger ---exponential--- 
in the case of CELs. The temperature fluctuations have been parametrized traditionally by {\ts}, 
the mean square temperature 
fluctuation of the gas \citep{peimbert67}. We have computed {\ts} values from the comparison of abundances derived 
using CELs or RLs, and 
from the temperatures derived from CELs and recombination processes. 

The main aims of this work are to present the high-quality spectrophotometric data 
for S~311 obtained with the ESO Very Large Telescope (VLT), to assess 
the old-fashioned abundance analysis of S~311 in the literature, calculate 
C$^{++}$ and O$^{++}$ abundances from recombination lines, and report 
the detection and measurement of weak deuterium Balmer lines.

In \S\S~\ref{obsred} and \ref{lin} we describe the observations, the data reduction procedure and 
the measurement and identification of the emission lines. 
In \S~\ref{phiscond} we obtain temperatures 
and densities using several diagnostic ratios. 
In \S~\ref{abundances} ionic abundances are 
determined based on CELs, as well as on RLs. In \S~\ref{tsq} we discuss the {\ts} results. 
Total abundances are determined in \S~\ref{totabun}. In \S~\ref{deut} we discuss the 
detection of deuterium Balmer lines. 
Finally, in \S\S~\ref{discus} and ~\ref{conclu} we
present the discussion and the conclusions, respectively.

\section{Observations and Data Reduction}\label{obsred}

\setcounter{table}{0}
\begin{table}
\begin{minipage}{75mm}
\centering \caption{Journal of observations.}
\label{tobs}
\begin{tabular}{c@{\hspace{2.8mm}}c@{\hspace{2.8mm}}c@{\hspace{2.8mm}}c@{\hspace{1.8mm}}}
\noalign{\hrule} \noalign{\vskip3pt}
Telescope& Date & $\Delta\lambda$~(\AA) & Exp. time (s)\\
\noalign{\vskip3pt} \noalign{\hrule} \noalign{\vskip3pt}
8.2 m VLT& 2003/03/30 & 3000$-$3900 & 3$\times$600 \\
"&" & 3800$-$5000 & 60,3$\times$1800 \\
"&" & 4750$-$6800 & 3$\times$600 \\
"&" & 6700$-$10400 & 60,3$\times$1800 \\
\noalign{\vskip3pt} \noalign{\hrule} \noalign{\vskip3pt}
\end{tabular}
\end{minipage}
\end{table}

The observations were made on 2003 March 30 with the Ultraviolet Visual Echelle Spectrograph, UVES 
\citep{dodoricoetal00}, 
at the VLT Kueyen Telescope in Cerro Paranal Observatory (Chile). We used the standard settings in 
both the red and 
blue arms of the spectrograph, covering the region from 3100 to 10400 \AA\ . The log of the 
observations is presented in Table~\ref{tobs}.

The wavelength regions 5783--5830 \AA\ and 8540--8650
\AA\ were not observed due to a gap between the two CCDs used in
the red arm. There are also five small gaps that were not observed, 9608--9612 \AA, 
9761--9767 \AA, 9918--9927 \AA, 10080--10093 \AA\ and 10249--10264 \AA, because  
the five redmost orders did not fit completely within the CCD.  We took long and short exposure 
spectra to check for possible saturation effects.

\begin{figure}
\begin{center}
\epsfig{file=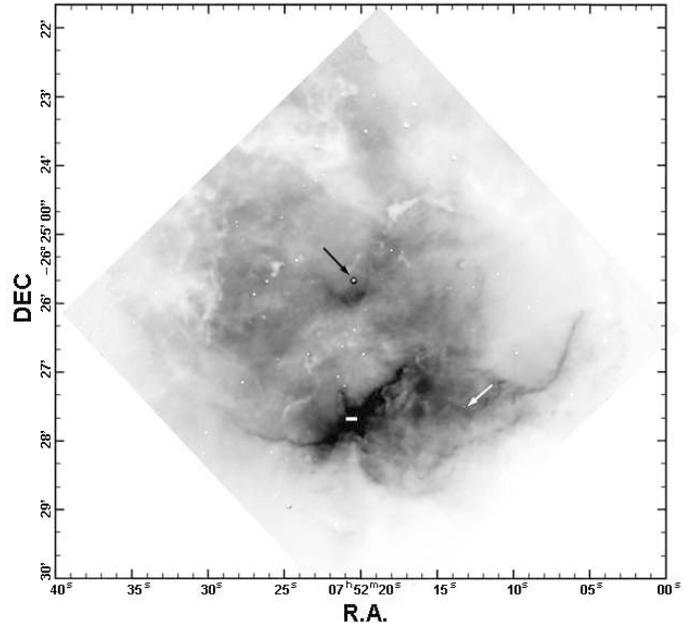, width=9. cm, clip=}
\caption{Deep continuum-subtracted H$\alpha$ image of S~311. The image was taken at 
the Nordic Optical Telescope (2.56m) in the Roque de los Muchachos Observatory, 
in La Palma (Spain) and is courtesy of A. R. L\'opez-S\'anchez. The black arrow 
indicates the position of the ionising star HD~64315. The white arrow 
indicates the position of the star NGC 2467--12 (see \S\ref{tsq})
The slit position is indicated as a white horizontal bar. 
\label{halpha}}
\end{center}
\end{figure}

The slit was oriented east-west and the atmospheric dispersion corrector (ADC) was used to keep 
the same observed
region within the slit regardless of the air mass value. The slit width was
set to 3.0$\arcsec$ and the slit length was set to 10$\arcsec$ in the blue arm and to 12$\arcsec$ 
in the red arm; the slit width was chosen to maximize the S/N ratio of the
emission lines and to maintain the required resolution to separate most of the
weak lines needed for this project. 
The effective resolution
at a given wavelength is approximately $\Delta \lambda \sim \lambda / 8800$. 
The centre of the slit was placed 126$\arcsec$ south of the main ionizing star 
HD 64315 (O6e), 
covering the brightest region of S~311 (see Figure~\ref{halpha}). The reductions were made for an area 
of 3$\arcsec$$\times$8.5$\arcsec$.

The spectra were reduced using the {\sc IRAF}\footnote{{\sc IRAF} is distributed by NOAO, 
which is operated by AURA, under cooperative agreement with NSF.} 
echelle reduction
package, following the standard procedure of bias subtraction, aperture
extraction, flatfielding, wavelength calibration and flux calibration. 
The standard stars EG 247, C-32d9927 \citep{hamuyetal92,hamuyetal94} and HD 49798 were 
observed for flux calibration.

\section{Line Intensities and Reddening Correction}\label{lin}

Line intensities were measured integrating all the flux in the line between two 
given limits and over a local continuum estimated by eye. In the cases of line blending, 
a multiple Gaussian profile fit procedure was applied to obtain the line flux of each 
individual line. Most of these measurements were made with the {\sc SPLOT} routine of the {\sc 
IRAF} 
package. In some cases of very tight blends or blends with very bright telluric lines the 
analysis was performed via Gaussian fitting making use of the Starlink {\sc DIPSO} software 
\citep{howardmurray90}.

\setcounter{table}{1}
\begin{table}
\begin{minipage}{75mm}
\centering \caption{Observed and reddening-corrected line ratios 
[F(H$\beta$)=100] and identifications.}
\label{lineid}
\begin{tabular}{l@{\hspace{2.8mm}}l@{\hspace{2.8mm}}c@{\hspace{1.8mm}}
c@{\hspace{2.8mm}}c@{\hspace{2.8mm}}c@{\hspace{2.8mm}}r@{\hspace{1.8mm}}l@{\hspace{1.8mm}}}
\noalign{\hrule} \noalign{\vskip3pt}
$\lambda_{\rm 0}$& & & $\lambda_{\rm obs}$& & & err & Notes \\
(${\rm \AA}$)& Ion& Mult.& (${\rm \AA}$)& $F(\lambda)$$^{\rm a}$& $I(\lambda)$$^{\rm b}$& (\%) & \\
\noalign{\vskip3pt} \noalign{\hrule} \noalign{\vskip3pt}
3187.84 &   He I   & 3       & 3188.45  &    2.383  &	 4.486  &   6 &  \\
3587.28 &   He I   & 32      & 3588.08  &    0.191  &	 0.294  &  20 &  \\
3613.64 &   He I   & 6       & 3614.45  &    0.276  &	 0.422  &  15 &  \\
3634.25 &   He I   & 28      & 3635.08  &    0.355  &	 0.537  &  12 &  \\
3656.11 &   H I    & H35     & 3658.75  &    0.070  &	 0.105  &  :  &  \\
3658.64 &   H I    & H34     & 3659.39  &    0.103  &	 0.154  &  34 &  \\
3659.42 &   H I    & H33     & 3660.26  &    0.135  &	 0.202  &  27 &  \\
3660.28 &   H I    & H32     & 3661.06  &    0.180  &	 0.269  &  21 &  \\
3661.22 &   H I    & H31     & 3662.10  &    0.234  &	 0.350  &  17 &  \\
3662.26 &   H I    & H30     & 3663.04  &    0.177  &	 0.264  &  22 &  \\
3663.40 &   H I    & H29     & 3664.25  &    0.210  &	 0.314  &  19 &  \\
3664.68 &   H I    & H28     & 3665.45  &    0.255  &	 0.380  &  16 &  \\
3666.10 &   H I    & H27     & 3666.91  &    0.256  &	 0.382  &  16 &  \\
3667.68 &   H I    & H26     & 3668.49  &    0.270  &	 0.402  &  15 &  \\
3669.47 &   H I    & H25     & 3670.26  &    0.315  &	 0.470  &  14 &  \\
3671.48 &   H I    & H24     & 3672.27  &    0.375  &	 0.559  &  12 &  \\
3673.76 &   H I    & H23     & 3674.57  &    0.381  &	 0.567  &  12 &  \\
3676.37 &   H I    & H22     & 3677.17  &    0.496  &	 0.739  &  10 &  \\
3679.36 &   H I    & H21     & 3680.16  &    0.486  &	 0.722  &  10 &  \\
3682.81 &   H I    & H20     & 3683.63  &    0.545  &	 0.808  &  9 &  \\
3686.83 &   H I    & H19     & 3687.63  &    0.617  &	 0.914  &  8 &  \\
3691.56 &   H I    & H18     & 3692.36  &    0.777  &	 1.150  &   7 &  \\
3697.15 &   H I    & H17     & 3697.97  &    0.900  &	 1.329  &   7 &  \\
3703.86 &   H I    & H16     & 3704.67  &    1.024  &	 1.504  &   6 &  \\
3705.04 &   He I   & 25      & 3705.84  &    0.407  &	 0.597  &  11 &  \\
3711.97 &   H I    & H15     & 3712.79  &    1.200  &	 1.756  &   6 &  \\
3721.83 &  [S III] & 2F      & 3722.68  &    2.156  &	 3.145  &   5 &  \\
3721.94 &  H I     & H14     & 	        &	    &  	        &     &  \\
3726.03 &  [OII]   & 1F      & 3726.88  &  113.36   &  165.15   &   4 &  \\
3728.82 &  [OII]   & 1F      & 3729.63  &  133.39   &  194.14   &   4 &  \\
3734.37 &   H I    & H13     & 3735.19  &    1.812  &	 2.633  &   5 &  \\
3750.15 &   H I    & H12     & 3750.97  &    2.305  &	 3.330  &   4 &  \\
3770.63 &   H I    & H11     & 3771.47  &    2.822  &	 4.048  &   4 &  \\
3797.90 &   H I    & H10     & 3798.74  &    3.721  &	 5.289  &   4 &  \\
3819.61 &   He I   & 20      & 3820.47  &    0.731  &	 1.031  &   7 &  \\
3831.66 &   S II   &         & 3832.51  &    0.026  &    0.036  &  29  & g \\
3833.57 &   He I   & 62      & 3834.37  &    0.117  &    0.165  &   10  & \\
3835.39 &   H I    & H9      & 3836.23  &    5.226  &	 7.338  &   4 &  \\
3856.02 &   Si II  & 1	     & 3856.97  &    0.046  &    0.064  &  20  & e \\
3856.13	&   O II   & 12	     &          &           &           &      &  \\
3867.48 &   He I   & 20      & 3868.31  &    0.084  &    0.117  &  12  &  \\
3868.75 & [Ne III] & 1F      & 3869.62  &    3.708  &	 5.152  &   4 &  \\
3871.60 &   He I   & 60      & 3872.67  &    0.076  &    0.105  &  14  &  \\
3888.65 &   He I   & 2       & 3889.50  &    5.177  &    7.149  &   4  & \\
3889.05 &   H I    & H8      & 3889.91  &    7.991  &   11.034  &   4  & \\
3920.68 &   C II   & 4       & 3921.47  &    0.034  &    0.046  &  25  &  \\
3926.53 &   He I   & 58      & 3927.44  &    0.098  &    0.134  &  11  &  \\
3964.73 &   He I   & 5       & 3965.61  &    0.668  &    0.902  &   4  &  \\
3967.46 & [Ne III] & 1F      & 3968.34  &    1.149  &    1.549  &   4  &  \\
3970.07 &   H I    & H7      & 3970.95  &   11.680  &   15.728  &   4  &  \\
4008.36 & [Fe III] & 4F      & 4009.25  &    0.029  &    0.039  &   28  & \\
4009.22 &   He I   & 55      & 4010.14  &    0.125  &    0.167  &   9  &  \\
4026.21 &   He I   & 18      & 4027.10  &    1.457  &    1.927  &   4  &  \\
4068.60 &  [S II]  & 1F      & 4069.51  &    1.710  &    2.227  &   4  &  \\
4076.35 &  [S II]  & 1F      & 4077.25  &    0.598  &    0.777  &   4  &  \\
4100.62 &    D I   & D6      & 4101.47  &    0.025  &    0.032  &  30  & \\
4101.74 &   H I    & H6      & 4102.64  &   19.255  &   24.792  &   3  & \\
4120.82 &  He I    & 16      & 4121.75  &    0.126  &    0.161  &   9  &  \\
4143.76 &   He I   & 53      & 4144.68  &    0.220  &    0.280  &  6  &  \\
4153.30 &  O II    & 19      & 4154.18  &    0.019  &    0.024  &  :  &  \\
\noalign{\vskip3pt} \noalign{\hrule} \noalign{\vskip3pt}
\end{tabular}
\end{minipage}
\end{table}

\setcounter{table}{1}
\begin{table}
\begin{minipage}{75mm}
\centering \caption{continued}
\begin{tabular}{l@{\hspace{2.8mm}}l@{\hspace{2.8mm}}c@{\hspace{1.8mm}}
c@{\hspace{2.8mm}}c@{\hspace{2.8mm}}c@{\hspace{2.8mm}}r@{\hspace{1.8mm}}l@{\hspace{1.8mm}}}
\noalign{\hrule} \noalign{\vskip3pt}
$\lambda_{\rm 0}$& & & $\lambda_{\rm obs}$& & & err & Notes \\
(${\rm \AA}$)& Ion& Mult.& (${\rm \AA}$)& $F(\lambda)$$^{\rm a}$& $I(\lambda)$$^{\rm b}$& (\%) & \\
\noalign{\vskip3pt} \noalign{\hrule} \noalign{\vskip3pt}
4168.97 &   He I   & 52      & 4169.90  &    0.035  &    0.044  &  24  & e \\
4169.22 &   O II   & 19      &          &	    &           &    &  \\       
4267.15 &   C II   & 6       & 4268.11  &    0.087  &    0.108  &  12  &  \\
4287.40 & [Fe II]  & 7F      & 4288.35  &    0.024  &    0.029  &  33  &  \\
4303.61 &   O II   & 66      & 4304.83  &    0.030  &    0.036  &  27  &  \\
4303.82 &   O II   & 53      &          &           &           &      &  \\
4339.29 &   D I    & D5      & 4340.26  &    0.036  &	 0.044  &  22  & \\
4340.47 &   H I    & H5      & 4341.42  &   38.644  &	46.737  &   3  & \\
4359.34 & [Fe II]  & 7F      & 4360.32  &    0.022  &    0.026  &  35  &  \\
4363.21 & [O III]  & 2F      & 4364.17  &    0.468  &    0.562  &   4  &  \\
4366.89 &   O II   & 2       & 4367.95  &    0.077  &    0.092  &  13  &  \\
4368.15 &   O I    & 5       & 4369.22  &    0.032  &    0.039  &  26  &  \\
4368.25 &   O I    & 5       &          &           &           &      &  \\
4387.93 &   He I   & 51      & 4388.90  &    0.439  &    0.522  &   4  &  \\
4437.55 &   He I   & 50      & 4438.52  &    0.063  &    0.073  &  15  &  \\
4471.48 &   He I   & 14      & 4472.49  &    3.521  &    4.056  &   3  &  \\
4562.60 &   Mg I]  & 1       & 4563.63  &    0.138  &    0.153  &   8  &  \\
4571.10 &   Mg I]  & 1       & 4572.13  &    0.120  &    0.134  &   9  &  \\
4630.54 &   N II   & 5       & 4631.65  &    0.019  &    0.021  &  39  &  \\
4638.86 &   O II   & 1       & 4639.88  &    0.026  &    0.028  &  30  &  \\
4641.81 &   O II   & 1       & 4642.85  &    0.026  &    0.028  &  30  &  \\
4649.13 &   O II   & 1       & 4650.16  &    0.023  &    0.025  &  33  &  \\
4650.84 &   O II   & 1       & 4651.81  &    0.023  &    0.025  &  33  &  \\
4654.12 &   O I    & 18      & 4655.05  &    0.013  &    0.014  &  :  &  \\
4658.10 & [Fe III] & 3F      & 4659.16  &    0.184  &    0.197  &   7  &  \\
4661.63 &   O II   & 1       & 4662.60  &    0.022  &    0.024  &  35  &  \\
4667.01 & [Fe III] & 3F      & 4667.91  &    0.021  &    0.022  &  :  &  \\
4701.62 & [Fe III] & 3F      & 4702.62  &    0.050  &    0.052  &  18  &  \\
4711.37 & [Ar IV]  & 1F      & 4712.51  &    0.012  &    0.012  &  :  &  e \\
4713.14 &    He I  & 12      & 4714.22  &    0.394  &    0.415  &   4  &  \\
4754.83 & [Fe III] & 3F      & 4755.80  &    0.041  &    0.042  &  21  &  \\
4769.60 & [Fe III] & 3F      & 4770.58  &    0.019  &    0.020  &  36  &  \\
4777.78 & [Fe III] & 3F      & 4778.76  &    0.015  &    0.016  &  :  &  \\
4814.55 & [Fe II]  & 20F     & 4815.64  &    0.018  &    0.018  &  :  &  \\
4815.51 &  S II    & 9	     & 4816.67  &    0.027  &    0.027  &  28 & \\
4860.03 &   D I    & D4      & 4861.08  &    0.085  &    0.085  &  18  & \\
4861.33 &   H I    & H4      & 4862.40  &   100.00  &   100.00  &   3  &  \\
4881.00 & [Fe III] & 2F      & 4882.13  &    0.059  &    0.059  &  16  &  \\
4921.93 &   He I   & 48      & 4923.03  &    1.117  &    1.093  &   4  &  \\
4924.50 & [Fe III] & 2F      & 4925.60  &    0.020  &    0.020  &  38  & \\
4924.50 & O II     & 28      &          &           &           &      & \\
4930.50 & [Fe III] & 1F      & 4931.63  &    0.017  &    0.016  &  :  &  \\
4931.32 & [O III]  & 1F      & 4932.28  &    0.028  &    0.027  &  29  &  \\
4958.91 & [O III]  & 1F      & 4960.02  &   44.315  &   42.916  &   3  &  \\
4985.90 & [Fe III] & 2F      & 4986.94  &    0.142  &    0.136  &  12  &  \\
5006.84 & [O III]  & 1F      & 5007.96  &  132.658  &  126.110  &   3  &  \\
5015.68 &   He I   & 4       & 5016.79  &    2.328  &    2.207  &   3  &  \\
5035.79 & [Fe II]  & 4F      & 5037.02  &    0.023  &    0.022  &   :  &  \\
5041.03 &   Si II  & 5       & 5042.25  &    0.027  &    0.025  &   :  &  \\
5047.74 &   He I   & 47      & 5048.92  &    0.225  &    0.211  &   10  &  \\
5055.98 &   Si II  & 5       & 5057.12  &    0.051  &    0.048  &  32  &  \\
5056.31 &   Si II  & 5       &		&	    &		&      &  \\ 	        
5191.82 & [Ar III] & 1F      & 5192.84  &    0.056  &    0.050  &  30  &  \\
5197.90 &  [N I]   & 1F      & 5199.12  &    0.389  &    0.349  &   7  &  \\
5200.26 &  [N I]   & 1F      & 5201.47  &    0.330  &    0.296  &   8  &  \\
5270.30 & [Fe III] & 1F      & 5271.70  &    0.124  &    0.108  &  16  &  \\
5517.71 & [Cl III] & 1F      & 5518.91  &    0.566  &    0.461  &   6  &  \\
5537.88 & [Cl III] & 1F      & 5539.06  &    0.437  &    0.354  &   7  &  \\
5577.34 &  [O I]   & 3F      & 5578.56  &    0.072  &    0.057  &  25  &  \\
5754.64 & [N II]   & 3F      & 5755.89  &    1.315  &    0.992  &   4  &  \\
\noalign{\vskip3pt} \noalign{\hrule} \noalign{\vskip3pt}
\end{tabular}
\end{minipage}
\end{table}

\setcounter{table}{1}
\begin{table}
\begin{minipage}{75mm}
\centering \caption{continued}
\begin{tabular}{l@{\hspace{2.8mm}}l@{\hspace{2.8mm}}c@{\hspace{1.8mm}}
c@{\hspace{2.8mm}}c@{\hspace{2.8mm}}c@{\hspace{2.8mm}}r@{\hspace{1.8mm}}l@{\hspace{1.8mm}}}
\noalign{\hrule} \noalign{\vskip3pt}
$\lambda_{\rm 0}$& & & $\lambda_{\rm obs}$& & & err & Notes \\
(${\rm \AA}$)& Ion& Mult.& (${\rm \AA}$)& $F(\lambda)$$^{\rm a}$& $I(\lambda)$$^{\rm b}$& (\%) & \\
\noalign{\vskip3pt} \noalign{\hrule} \noalign{\vskip3pt}
5875.64 &   He I   & 11      & 5876.96  &   15.602  &   11.343  &   4  &  \\
5978.83 & Si II    & 4       & 5980.31  &    0.076  &    0.054  &   24 &  \\
6046.44 &   O I    & 22      & 6047.77  &    0.034  &    0.024  &   :  &  \\
6300.30 &  [O I]   & 1F      & 6301.75  &    3.512  &    2.307  &   4  &  \\
6312.10 & [S III]  & 3F      & 6313.48  &    2.057  &    1.348  &   4  &  \\
6363.78 &  [O I]   & 1F      & 6365.24  &    1.258  &    0.815  &   5  &  \\
6371.36 & Si II    &  2      & 6372.85  &    0.064  &    0.041  &   27 &  \\
6548.03 &  [N II]  & 1F      & 6549.54  &   42.523  &   26.484  &   4  &  \\
6561.04 &   D I    & D3      & 6562.43  &    0.321  &	 0.200  &   9  & \\
6562.82 &   H I    & H3      & 6564.26  &  458.723  &  284.845  &   4  & \\
6578.05 &   C II   & 2       & 6579.49  &    0.196  &    0.121  &   12  &  \\
6583.41 &  [N II]  & 1F      & 6584.93  &  131.646  &   81.408  &   4  &  \\
6678.15 &   He I   & 46      & 6679.64  &    5.374  &    3.263  &   4  &  \\
6716.47 &  [S II]  & 2F      & 6717.97  &   43.847  &   26.434  &   4  &  \\
6730.85 &  [S II]  & 2F      & 6732.35  &   39.503  &   23.752  &   4  &  \\
6933.91 &   He I   & 1/13    & 6935.57  &    0.017  &    0.010  &  :  &  g \\
6989.45 &  He I    & 1/12    & 6991.12  &    0.024  &    0.014  &   :  &  \\
7002.23 &  O I     & 21      & 7003.53  &    0.179  &    0.103  &   9  &  c \\
7065.28 &   He I   & 10      & 7066.81  &    3.780  &    2.149  &   5  &  \\
7092.19 & C I]     &         & 7093.57  &    0.039  &    0.022  &  32  &  \\
7093.97 & [Fe II]  &         & 7095.54  &    0.013  &    0.007  &  :  & g \\
7105.85 &   Si I   & 70      & 7107.13  &    0.029  &	0.016  &  :  & g \\
7111.47 &   C I	   &	     & 7113.06  &    0.034  &    0.019  &  37  &  g \\
7135.78 & [Ar III] & 1F      & 7137.37  &   17.966  &   10.102  &   5  &  \\
7155.14 & [Fe II]  & 14F     & 7156.81  &    0.029  &    0.016  &  :  &  \\
7160.58 &  He I    & 1/10    & 7162.20  &    0.048  &    0.027  &  27  &  \\
7231.34 &  C II    & 3       & 7232.89  &    0.070  &    0.039  &  19  &  \\
7236.19 &  C II    & 3       & ---      &    ---    &    ---    &   -   & c \\
7281.35 &  He I    & 45      & 7282.98  &    1.055  &    0.580  &   5  &  \\
7298.05 &  He I    & 1/9     & 7299.71  &    0.039  &    0.021  &  33  &  \\
7318.39 &  [O II]   & 2F      & 7320.69  &	1.985  &    1.085  &   4  &  \\
7319.99 &  [O II]   & 2F      & 7321.77  &	6.024  &    3.293  &   5  &  \\
7329.66 &  [O II]   & 2F      & 7331.33  &	3.311  &    1.807  &   5  &  \\
7330.73 &  [O II]   & 2F      & 7332.41  &	3.133  &    1.710  &   5  &  \\
7377.83 &  [Ni II] & 2F      & 7379.55  &    0.019  &    0.010  &  :  &  \\
7423.64 &  N I     & 3       & 7425.40  &    0.026  &    0.014  &  :  &  \\
7442.30 &  N I     & 3       & 7444.02  &    0.050  &    0.027  &  26  &  \\
7468.31 &  N I     & 3       & 7470.14  &    0.156  &    0.083  &  10  &  \\
7499.85 &  He I    & 1/8     & 7501.53  &    0.066  &    0.035  &  20  &  \\
7706.74 &  O I     & 42      & 7708.80  &    0.036  &    0.019  &  35  &   \\
7751.10 & [Ar III] & 2F      & 7752.84  &    4.854  &    2.500  &   5  &  \\
7782.10 &  Ca I]   &        & 7784.06  &    0.073  &    0.038  &  18  &   \\
7790.60	& Ar I	   &	     & 7792.83  &    0.064  &    0.033  &  21  &  \\
7801.56 & [Cr II]  &       & 7803.60  &    0.051  &    0.026  &  25  &  \\
7816.13 & He I     & 1/7     & 7817.90  &    0.106  &    0.054  &  14  &  \\
7837.85 & [Co I]   &       & 7839.58  &    0.105  &    0.053  &  14  & \\
7862.75 & Fe II]    &       & 7864.55  &    0.023  &    0.012  &  :  & g \\
7875.99 & [P II]   &$^1$D-$^1$S& 7877.68  &    0.066  &    0.033  &  20  &   \\
7959.70 & Co II]   &	   & 7961.62  &    0.109  &   0.055   &  13  & g \\
8046.80 & Si I     & 73      & 8048.32  &    0.048  &    0.024  &  27  & g \\
8116    & He I     & 4/16    & 8116.93  &    0.008  &    0.004  &  :  &  \\
8150.57 & Si I     & 20      & 8152.52  &    0.077  &    0.038  &  19  &   \\
8184.85 & N I      & 2       & 8186.78  &    0.029  &    0.014  &  :  &  \\
8188.01 & N I      & 2       & 8189.93  &    0.078  &    0.038  &  18  &  \\
8200.36 & N I      & 2       & 8202.33  &    0.017  &    0.008  &  :  &  \\
8210.72 & N I      & 2       & 8212.65  &    0.036  &    0.017  &  35  &  \\
8216.28 & N I      & 2       & 8218.25  &    0.085  &    0.042  &  16  &  \\
8223.14 & N I      & 2       & 8225.03  &    0.095  &    0.046  &  15  &  \\
8243.70 & H I      & P43     & 8245.54  &    0.068  &    0.033  &  20  &  \\
8245.64 & H I      & P42     & 8247.66  &    0.094  &    0.045  &  15  &  \\
\noalign{\vskip3pt} \noalign{\hrule} \noalign{\vskip3pt}
\end{tabular}
\end{minipage}
\end{table}

\setcounter{table}{1}
\begin{table}
\begin{minipage}{75mm}
\centering \caption{continued}
\begin{tabular}{l@{\hspace{2.8mm}}l@{\hspace{2.8mm}}c@{\hspace{1.8mm}}
c@{\hspace{2.8mm}}c@{\hspace{2.8mm}}c@{\hspace{2.8mm}}r@{\hspace{1.8mm}}l@{\hspace{1.8mm}}}
\noalign{\hrule} \noalign{\vskip3pt}
$\lambda_{\rm 0}$& & & $\lambda_{\rm obs}$& & & err & Notes \\
(${\rm \AA}$)& Ion& Mult.& (${\rm \AA}$)& $F(\lambda)$$^{\rm a}$& $I(\lambda)$$^{\rm b}$& (\%) & \\
\noalign{\vskip3pt} \noalign{\hrule} \noalign{\vskip3pt}
8247.73 & H I      & P41     & 8249.60  &    0.090  &    0.044  &  16  &  \\
8249.97 & H I      & P40     & 8251.80  &    0.081  &    0.039  &  17  &  \\
8252.40 & H I      & P39     & 8254.35  &    0.130  &    0.063  &  12  &  \\
8255.02 & H I      & P38     & 8256.88  &    0.104  &    0.050  &  14  &  \\
8257.85 & H I      & P37     & 8259.72  &    0.104  &    0.050  &  14  &  \\
8260.93 & H I      & P36     & 8262.76  &    0.112  &    0.054  &  13  &  \\
8264.28 & H I      & P35     & 8266.18  &    0.159  &    0.077  &  10  & \\
8266.40 & Ar I     &         & 8268.32  &    0.092  &    0.045  &  15  &  \\
8267.94 & H I      & P34     & 8269.78  &    0.152  &    0.074  &  10  &  \\
8271.93 & H I	   & P33     & 8273.63  &    0.086  &    0.041  &  16  & d \\
8276.31 & H I	   & P32     & ---     &    ---   &    ---   &  --  & c \\
8281.12 & H I	   & P31     & 8282.85  &    0.270  &    0.130  &   8  & c \\
8286.43 & H I      & P30     & ---     &    ---   &    ---   &  --  & c \\
8292.31 & H I	   & P29     & 8294.22  &    0.177  &    0.085  &  9  &  \\
8298.83 & H I	   & P28     & 8300.72  &    0.252  &    0.121  &   8  &  \\
8306.11 & H I	   & P27     & 8307.94  &    0.286  &    0.138  &   7  &  \\
8314.26 & H I	   & P26     & 8316.18  &    0.301  &    0.145  &   7  &  \\
8323.42 & H I	   & P25     & 8325.28  &    0.339  &    0.162  &   7  &  \\
8333.78 & H I	   & P24     & 8335.63  &    0.434  &    0.207  &   6  &  \\
8334.75 & Fe II]   &         & 8336.68  &    0.190  &    0.091  &   9  & g \\
8345.55 & H I	   & P23     & 8347.42  &    0.423  &    0.202  &   6  &  \\
8359.00 & H I	   & P22     & 8360.80  &    0.629  &    0.299  &   6  &  \\
8374.48 & H I	   & P21     & 8376.33  &    0.549  &    0.260  &   6  &  \\
8387.77 & Fe I     &         & 8389.62  &    0.095  &    0.045  &  15  & g \\
8392.40 & H I	   & P20     & 8394.24  &    0.634  &    0.299  &   6  &  \\
8395.98 & Ca I]   &         & 8397.84  &    0.059  &    0.028  &  23  & c, g \\
8413.32 & H I	   & P19     & 8415.20  &    0.713  &    0.336  &   6  & c \\
8437.96 & H I	   & P18     & 8439.85  &    0.818  &    0.383  &   6  &  \\
8446.25 & O I      & 4       & 8448.41  &    0.823  &    0.384  &   6  &  \\
8446.36 & O I      & 4       &          &           &           &      &  \\
8446.76 & O I      & 4       & 8448.93  &    0.056  &    0.026  &  24  &  \\
8451.00 & He I     & 6/17    & 8453.21  &    0.043  &    0.020  &  30  &  \\
8459.32 & [Cr II]  &       & 8461.26  &    0.193  &    0.090  &   9  & \\
8467.25 & H I      & P17     & 8469.15  &    0.936  &    0.435  &   6  &  \\
        & ?        &         & 8477.02  &    0.032  &    0.015  &  39  &  \\
8486.   & He I      & 6/16    & 8488.26  &    0.034  &    0.016  &  37  &  \\
8502.48 & H I      & P16     & 8504.33  &    3.337  &    1.538  &   6  & c \\
8518.04 & He I     & 2/8     & 8520.05  &    0.076  &    0.035  &  18  &  \\
8528.99 & He I     & 6/15    & 8530.87  &    0.060  &    0.027  &  22  & c \\
8665.02 & H I      & P13     & 8666.95  &    2.136  &    0.947  &   6  &  \\
8680.28 & N I      &  1      & 8682.31  &    0.064  &    0.028  &  21  &  \\
8683.40 & N I      &  1      & 8685.38  &    0.152  &    0.067  &  11  &  \\
8686.15 & N I      &  1      & 8688.20  &    0.044  &    0.020  &  29  &  \\
8703.25 & N I      &  1      & 8705.22  &    0.051  &    0.022  &  26  & c \\
8711.70 & N I      &  1      & 8713.75  &    0.048  &    0.021  &  27  &  \\
8718.83 & N I      &  1      & 8720.90  &    0.029  &    0.013  &  :  &  \\
8727.13 & [C I]    &       & 8729.19  &    0.470  &    0.205  &   7  &  \\
8733.43 & He I     & 6/12    & 8735.41  &    0.074  &    0.032  &  18  &  \\
8750.47 & H I      & P12     & 8752.44  &    2.672  &    1.161  &   6  &  \\
8776.77 & He I     & 4/9     & 8778.70  &    0.074  &    0.032  &  18  &  \\
8845.38 & He I     & 6/11    & 8847.36  &    0.100  &    0.043  &  15  &  \\
8850.62 & Fe I]    &       & 8852.74  &    0.142  &    0.060  &  11  &  g \\
8862.79 & H I      & P11   & 8864.76  &    3.597  &    1.525  &   6  &  \\
8888.71 & Fe I]    &       & 8890.93  &    0.082  &    0.035  &  17  & g \\
8893.87 & V I]    &       & 8895.90  &    0.203  &    0.086  &  9 & c,g \\
8914.77 & He I     & 2/7     & 8916.58  &    0.062  &    0.026  &  22  &  \\
8946.05 & Fe II]   &       & 8948.25  &    0.088  &    0.037  &  16  & g \\
8996.99 & He I     & 6/10    & 8998.95  &    0.146  &    0.060  &  12  &  \\
9014.91 & H I      & P10     & 9016.97  &    4.278  &    1.764  &   6  & d \\
9019.14 & Fe I]    &       & 9021.38  &    0.235  &    0.097  &   9  & g \\
\noalign{\vskip3pt} \noalign{\hrule} \noalign{\vskip3pt}
\end{tabular}
\end{minipage}
\end{table}

\setcounter{table}{1}
\begin{table}
\begin{minipage}{75mm}
\centering \caption{continued}
\begin{tabular}{l@{\hspace{2.8mm}}l@{\hspace{2.8mm}}c@{\hspace{1.8mm}}
c@{\hspace{2.8mm}}c@{\hspace{2.8mm}}c@{\hspace{2.8mm}}r@{\hspace{1.8mm}}l@{\hspace{1.8mm}}}
\noalign{\hrule} \noalign{\vskip3pt}
$\lambda_{\rm 0}$& & & $\lambda_{\rm obs}$& & & err & Notes \\
(${\rm \AA}$)& Ion& Mult.& (${\rm \AA}$)& $F(\lambda)$$^{\rm a}$& $I(\lambda)$$^{\rm b}$& (\%) & \\
\noalign{\vskip3pt} \noalign{\hrule} \noalign{\vskip3pt}
9019.14 & Ca I]    &       &          &           &           &      & \\
9029.07 & Fe I]    &       & 9031.34  &    0.145  &    0.060  &   11  & g  \\
9063.29 & He I     & 4/8     & 9065.18  &    0.222  &    0.091  &   9  &  \\
9068.90 & [S III]  & 1F      & 9070.92  &   55.891  &   22.855  &   6  &  \\
9094.83 & C I      &       & 9096.98  &    0.139  &    0.056  &  12  & g \\
9111.83 & Ca I]    &       & 9113.95  &    0.048  &    0.020  &  27  & g \\
9113.60 & Fe I]    &       & 9115.67  &    0.108  &    0.044  &  14  & g \\
9123.60 & [Cl II]  &       & 9125.72  &    0.225  &    0.091  &   9  &  \\
9162.65 & Ni II]   &       & 9164.72  &    0.047  &    0.019  &  28  & g \\
9210.28 & He I     & 6/9       & 9212.41  &    0.189  &    0.076  &  10  &  \\
9226.62 & [Fe II]  &       & 9228.58  &    0.088  &    0.036  &  16  & g \\
9229.01 & H I      & P9        & 9231.11  &    6.898  &    2.770  &   6  &  \\
9463.57 & He I     & 1/5       & 9465.71  &    0.286  &    0.113  &   8  &  \\
9504.54 & C I]     &       & 9506.97  &    0.127  &    0.050  &  12  & g \\
9526.16 & He I     & 6/8   & 9528.46  &    0.362  &    0.143  &   8  &  \\
9530.60 & [S III]  & 1F      & 9533.12  &  156.202  &   61.749  &   5  &  \\
9545.97 & H I      & P8      & 9548.13  &    7.956  &    3.144  &   7  & d \\
9702.62 & He I     & 75      & 9705.23  &    0.130  &    0.051  &  12  & c\\
9824.13 & [C I]    & 1F      & 9826.49  &    0.986  &    0.389  &   7  &  \\
9850.24 & [C I]    & 1F      & 9852.56  &    2.766  &    1.091  &   7  &  \\
9876.67 & Fe I]    &        & 9879.00  &    0.233  &    0.092  &   9  &  g \\
10027.70 & He I    & 6/7     &10029.98  &    0.595  &    0.235  &   7  &  \\
 10049.37 & H I     & P7      &10051.64  &   17.935  &    7.086  &   7  &  \\
10286.70 & [S II]   & 3F      &10288.95  &    1.978  &    0.782  &   7  &  \\
10320.49 & [S II]   & 3F      &10322.75  &    2.235  &    0.884  &   7  &  \\
10336.41 & [S II]   & 3F      &10338.96  &    3.066  &    1.212  &   7  &  \\
\noalign{\vskip3pt} \noalign{\hrule} \noalign{\vskip3pt}
\end{tabular}
\begin{description}
\item[$^{\rm a}$] Where $F$ is the observed flux in units of $100.00=3.703 \times 10^{-13}$ ergs 
cm$^{-2}$ s$^{-1}$.
\item[$^{\rm b}$] Where $I$ is the dereddened flux, in units of 
$100.00=1.616 \times 10^{-12}$ ergs cm$^{-2}$ s$^{-1}$ and assuming C(H$\beta$)=0.64 dex.
\item[$^{\rm c}$] Affected by telluric emission lines.
\item[$^{\rm d}$] Affected by atmospheric absorption bands.
\item[$^{\rm e}$] Affected by internal reflections or charge transfer in the CCD.
\item[$^{\rm f}$] Blend with an unknown line.
\item[$^{\rm g}$] Dubious identification.
\end{description}\end{minipage}
\end{table}

Table~\ref{lineid} presents the emission line intensities of S~311. The
first and fourth columns list the adopted laboratory wavelength, $\lambda_0$, and 
the observed wavelength in the heliocentric framework, $\lambda$. 
The second and third columns give the ion and the multiplet number, or 
series for each line.  The fifth and sixth columns give the observed
flux relative to H$\beta$, $F(\lambda$), and the flux corrected for reddening
relative to H$\beta$, $I(\lambda$). The seventh column gives the
fractional error (1$\sigma$) in the line intensities \citep[see][for details in error 
analysis]{garciarojasetal04}.

A total of 263 emission lines were measured; of them 178 are permitted, 65 are forbidden and 19 
are semiforbidden (see Table~\ref{lineid}). Two {\hi} Paschen lines and one {\cii} line
are blended with telluric lines, making impossible their measurement. Several 
other lines were strongly affected by atmospheric features in absorption, by internal 
reflections or charge transfer in the CCD, rendering their intensities unreliable. Also, 23 lines 
are dubious identifications and one emission line could not be identified in any of the available 
references. Those lines are indicated in Table~\ref{lineid}.

The identification and adopted laboratory wavelengths of the lines were obtained following 
several previous identifications in the literature \citep[see][and references therein]{garciarojasetal04, 
estebanetal04}

We have assumed the standard extinction for the Milky Way (R$_v$=3.1) parametrized by 
\citet{seaton79}. 
We have derived a logarithmic interstellar extinction coefficient of $c(H\beta)$=0.64 $\pm$ 0.04 
dex was determined by fitting the observed 
$I$(H Balmer lines)/$I$(H$\beta$) ratios (from H16 to H$\beta$) and $I$(H Paschen 
lines)/$I$(H$\beta$) (from P22 to P7), to the theoretical 
ones computed by Storey \& Hummer (1995) for {\te} = 10000 K and {\elecd} = 1000 cm$^{-3}$ (see 
below). {\hi} lines affected by blends or atmospheric absorption were not considered. 
The derived value of $c(H\beta)$ is in very good agreement with previous determinations in 
the same nebula: \citet{hawley78} derived $c(H\beta)$=0.61 and 0.67 for two slit positions with 
offsets of 96$\arcsec$ north and 96$\arcsec$ north, 35$\arcsec$ west respectively 
from our slit position. Furthermore, \citet{peimbertetal78} derived $c(H\beta)$=0.6 and 0.7 for 
slit positions 33$\arcsec$ north, 12$\arcsec$ west and 33$\arcsec$ north, 106$\arcsec$ west, 
respectively. These last authors used the \citet{whitford58} extinction law, which is almost coincident with 
the one we have adopted. Moreover, \citet{kennicuttetal00}, 
derived a value of $c(H\beta)$=0.67 using the average interstellar reddening curve from 
\citet{cardellietal89}. 
\citet{shaveretal83} observed two positions in S~311, one of them (position 2), almost coincident 
with our slit position. They derived $c(H\beta)$=0.5 and 0.7 for their positions 1 and 2 respectively.
We can conclude then that apparently there are no significant variations of the extinction 
inside S~311.

\section{Physical Conditions}\label{phiscond}

The large number of emission lines identified and measured in the spectra allows us the derivation 
of physical conditions using different emission
line ratios. The temperatures and densities are presented in Table~\ref{plasma}. Most of the 
determinations were carried out with the 
{\sc IRAF} task {\sc TEMDEN} of the package {\sc NEBULAR} \citep{shawdufour95}. 

The methodology followed for the derivation of {\elecd} and {\te} has been described in a previous 
paper 
\citep[i.e.][]{garciarojasetal04}. In the case of electron densities, ratios of CELs of several ions have been 
used.
The lastest version of {\sc NEBULAR} (February 2004) uses the transition probabilities recommended by 
\citet{wieseetal96} and the collision strengths of \citet{mclaughlinbell93} for the {\foii} 
$\lambda$3729/$\lambda$3726 doublet ratio. 
These atomic data yield electron densities systematically lower than those deduced from the 
{\fsii} $\lambda$6716/$\lambda$6731 doublet ratio \citep[see][]{estebanetal04, garciarojasetal04}. 
Following the 
arguments of \citet{copettiwritzl02} and \citet{wangetal04} we have adopted 
the transition probabilities from \citet{zeippen82} and the collision strengths from \citet{pradhan76} for 
the 
{\foii} $\lambda$3729/$\lambda$3726 doublet ratio, which give electron densities that are in good
agreement with the other density indicators.
We have derived the {\ffeiii} density from the intensity of the 6 brightest 
lines lines, which have errors less than 30 \% and seem not to be affected by line 
blending, 
together with the computations of \citet{rodriguez02}. All the computed values of {\elecd} are consistent 
within the errors (see Table~\ref{plasma}).

\setcounter{table}{2}
\begin{table}
\begin{minipage}{90mm}
\centering \caption{Plasma Diagnostic.}
\label{plasma}
\begin{tabular}{lll}
\noalign{\hrule} \noalign{\vskip3pt}
Parameter & Line & Value \\
\noalign{\vskip3pt} \noalign{\hrule} \noalign{\vskip3pt}
$N_{\rm e}$ (cm$^{-3}$)& {\fni} ($\lambda$5198)/($\lambda$5200)& 590$^{+260}_{-200}$  \\
& {\foii} ($\lambda$3726)/($\lambda$3729)& 260  $\pm$  110 \\
& {\fsii} ($\lambda$6716)/($\lambda$6731)& 360$^{+140}_{-120}$ \\ 
& {\ffeiii} &  390 $\pm$ 220\\
& {\fcliii} ($\lambda$5518)/($\lambda$5538)& 550$^{+650}_{-550}$ \\ 
& $N_{\rm e}$ (adopted) & 310 $\pm$ 80 \\
& & \\
$T_{\rm e}$ (K)& {\fnii} ($\lambda$6548+$\lambda$6583)/($\lambda$5755)& 9500  $\pm$  250$^{\rm a}$ 
\\
& {\fsii} ($\lambda$6716+$\lambda$6731)/($\lambda$4069+$\lambda$4076)&  7200 $^{+750}_{-600}$  \\
& {\foii} ($\lambda$3726+$\lambda$3729)/($\lambda$7320+$\lambda$7330)& 9800 $\pm$ 600$^{\rm a}$ \\
& $T_{\rm e}$ (low) & 9550 $\pm$ 250 \\
& {\foiii} ($\lambda$4959+$\lambda$5007)/($\lambda$4363) & 9000  $\pm$  200 \\
& {\fariii} ($\lambda$7136+$\lambda$7751)/($\lambda$5192)& 8800 $^{+700}_{-850}$ \\
& {\fsiii} ($\lambda$9069+$\lambda$9532)/($\lambda$6312) & 9300  $\pm$ 350 \\
& $T_{\rm e}$ (high) & 9050 $\pm$ 200 \\
& {\hei} & 8750 $\pm$ 500 \\
& Balmer line/cont. & 9500 $\pm$ 900 \\ 
& Paschen line/cont. & 8700 $\pm$ 1100 \\ 
\noalign{\vskip3pt} \noalign{\hrule} \noalign{\vskip3pt}
\end{tabular}
\begin{description}
\item[$^{\rm a}$] Recombination contribution on the auroral lines has been taken into account (see 
text)
\end{description}\end{minipage}
\end{table}

A weighted mean of {\elecd}({\oii}), {\elecd}({\feiii}), {\elecd}({\cliii}) and 
{\elecd}({\sii}) has been used to derive {\te}({\nii}), 
{\te}({\oiii}), {\te}({\ariii}) and {\te}({\siii}), and iterated until convergence. So, 
for all the species we have adopted {\elecd}=310 $\pm$ 80 cm$^{-3}$. 
We have excluded {\elecd}({\nitroi}) from the average because this ion is representative of the 
very outer part of the nebula, and probably does not coexist with most of the other ions.

Electron temperatures have been derived from the ratio of CELs of several ions and making use of 
{\sc NEBULAR} routines. In the case of {\te}({\siii}) instead of using the collision strengths of
\citet{galavisetal95}, included by default in {\sc NEBULAR}, we have used the ones by \citet{tayalgupta99}. The last
set of collision strengths gives {\te}({\siii}) values more consistent with the rest of temperature
determinations. Table~\ref{atom} shows the atomic data that we have changed in the last version of 
{\sc NEBULAR}.

\setcounter{table}{3}
\begin{table}
\begin{minipage}{75mm}
\centering \caption{Atomic data used for some selected ions.}
\label{atom}
\begin{tabular}{lcc}
\noalign{\hrule} \noalign{\vskip3pt}
Ion & Coll. Strengths & Trans. Probability \\
\noalign{\vskip3pt} \noalign{\hrule} \noalign{\vskip3pt}
O$^{+}$& \citet{pradhan76}& \citet{zeippen82}  \\
S$^{+}$&\citet{keenanetal96} & \citet{keenanetal93} \\
S$^{++}$& \citet{tayalgupta99} & \citet{mendozazeippen82b} \\
\noalign{\vskip3pt} \noalign{\hrule} \noalign{\vskip3pt}
\end{tabular}
\end{minipage}
\end{table}

To obtain  {\te}({\oii}) it is necessary to subtract the contribution to 
$\lambda\lambda$7320+7330 due to recombination;
 \citet{liuetal00} find that the contribution to the 
intensities of the {\foii} $\lambda\lambda$ 
7319, 7320, 7331, and 7332 lines due to recombination can be fitted in the range 
0.5$\le$T/10$^4$$\le$1.0 by:
\begin{equation}
\frac{I_R(7320+7330)}{I({\rm H\beta})}
= 9.36\times(T_4)^{0.44} \times \frac{{\rm{O}}^{++}}{{\rm{H}}^+},
\end{equation}
where $T_4$=$T$/10$^4$. With this equation we estimate a contribution of 
approximately 2\% to the observed line intensities.

\citet{liuetal00} also determined that the contribution to the intensity of the $\lambda$ 5755 
{\fnii} line due to recombination can be estimated from:
\begin{equation}
\frac{I_R(5755)}{I({\rm H\beta})}
= 3.19\times(T_4)^{0.30} \times \frac{{\rm N}^{++}}{{\rm H}^+},
\end{equation}
in the range 0.5$\le$ $T$/10$^4$$\le$2.0. We have obtained a contribution of recombination of
about 0.5\%, that does not affect significantly the temperature determination. 

\begin{figure}
\begin{center}
\epsfig{file=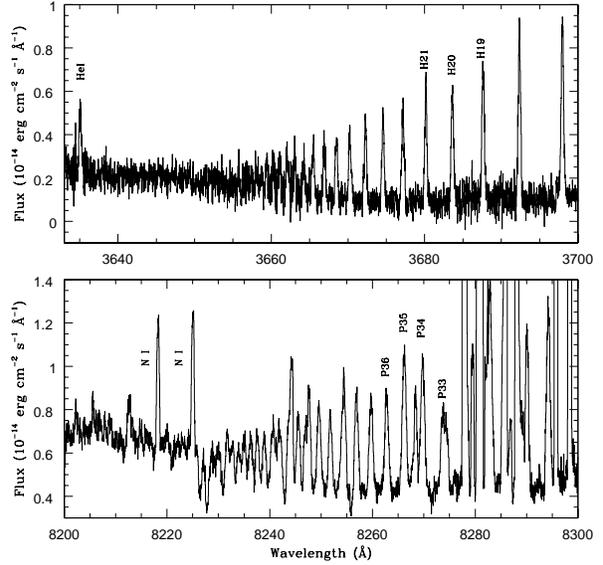, width=8. cm, clip=}
\caption{Section of the echelle spectrum including the Balmer (top) and the Paschen (bottom) 
limits (observed fluxes).
\label{saltos}}
\end{center}
\end{figure}

Figure~\ref{saltos} shows the spectral regions near the Balmer and the Paschen limits. The 
discontinuities can be easily appreciated. They are defined as 
$I_c(Bac)=I_c(\lambda3646^-)-I_c(\lambda3646^+)$ and 
$I_c(Pac)=I_c(\lambda8203^-)-I_c(\lambda8203^+)$ respectively. The high spectral 
resolution of the spectra permits to measure the continuum emission in zones very near de 
discontinuity, minimizing the possible contamination of other continuum 
contributions. We have obtained power-law fits to the relation between 
$I_c(Bac)/I(Hn)$ or $I_c(Pac)/I(Pn)$ and {\te} for different $n$ corresponding to 
different observed lines of both series. The emissivities as a function of electron 
temperature for the nebular continuum and the {\hi} Balmer and Paschen lines 
have been taken from \citet{brownmathews70} and \citet{storeyhummer95} respectively. The $T_e(Bac)$ 
adopted is the average of the values using the lines from $H\alpha$ to 
H~10 (the brightest ones). In the case of $T_e(Pac)$, the adopted value is the average 
of the individual temperatures obtained using the lines from P~7 to P~13 (the brightest 
lines of the series), excluding P~8 and P~10 because their intensity seems to be 
affected by sky absorption. 

\citet{apeimbertetal02} developed a method to derive the helium temperature, {\te}({\hei}), in the 
presence of temperature fluctuations. Assuming a 2-zone ionization scheme and the formulation 
of \citet{apeimbertetal02} we have derived {\te}({\hei})=8750 $\pm$ 500 K, which is highly 
consistent with {\te}({\hi}) assumed above.

We have assumed a 2-zone ionization scheme for the derivation of ionic abundances 
(see \S~\ref{abundances}). We have adopted the average of electron temperatures 
obtained from {\fnii} and {\foii} lines as representative for the low ionization zone, 
and the average of the values obtained from {\foiii}, [{\ariii}] and {\fsiii} lines for 
the high ionization zone (see Table~\ref{plasma}).

\section{Ionic Abundances}\label{abundances}

\subsection{He$^{+}$ abundance}\label{heabund}

\setcounter{table}{4}
\begin{table}
\begin{minipage}{75mm}
\centering \caption{He$^+$ abundance.}
\label{he}
\begin{tabular}{lcc}
\noalign{\hrule} \noalign{\vskip3pt}
Line & He$^+$/H$^+$ $^{\rm a}$\\
\noalign{\vskip3pt} \noalign{\hrule} \noalign{\vskip3pt}
3819.61& 775 $\pm$ 54 \\
3888.65& 759 $\pm$ 30 \\
3964.73& 825 $\pm$ 33\\ 
4026.21& 818 $\pm$ 33\\ 
4387.93& 837 $\pm$ 34\\ 
4471.09& 792 $\pm$ 24 \\
4713.14& 821 $\pm$ 33 \\
4921.93& 800 $\pm$ 32\\ 
5875.64& 780 $\pm$ 31\\ 
6678.15& 793 $\pm$ 32 \\
7065.28& 770 $\pm$ 38\\ 
7281.35& 835 $\pm$ 42\\ \hline
Adopted& 795 $\pm$ 9$^{\rm b}$\\ 
\noalign{\vskip3pt} \noalign{\hrule} \noalign{\vskip3pt}
\end{tabular}
\begin{description}
\item[$^{\rm a}$] In units of 10$^{-4}$, for $\tau_{3889}$=2.52 $\pm$ 0.44, and 
{\ts}=0.034 $\pm$ 0.010.
Uncertainties correspond to line intensity errors.
\item[$^{\rm b}$] It includes all the relevant uncertainties in emission line intensities, 
{\elecd}, $\tau_{3889}$ and {\ts}.
\end{description}
\end{minipage}
\end{table}

We have measured 50 {\hei} emission lines identified in our spectra. These lines arise mainly from 
recombination but they can be affected by collisional excitation and self-absorption effects. We 
have
determined the He$^+$/H$^+$ ratio using the effective recombination coefficients of \citet{storeyhummer95} 
for
{\hi} and those of \citet{smits96} and \citet{benjaminetal99} for {\hei}. The collisional contribution
was estimated from \citet{saweyberrington93} and \citet{kingdonferland95}, and the optical depths in the triplet lines 
were
derived from the computations by \citet{benjaminetal02}. From a maximum likelihood method 
\citep[e.g.][]{peimbertetal00},
using {\elecd}=310 $\pm$ 80 cm$^{-3}$ and $T$({\oii}+{\sc iii})=9600 $\pm$ 450 K (see \S~\ref{tsq}), we have
obtained He$^+$/H$^+$=0.0795 $\pm$ 0.0009, $\tau_{3889}$=2.52 $\pm$ 0.44, and {\ts}=0.034 $\pm$ 0.010. 
In Table~\ref{he} we have included the He$^+$/H$^+$ ratios we have obtained for the individual 
{\hei} lines not affected by line blending and with the highest signal-to-noise ratio. We have
excluded {\hei} $\lambda$5015 for the same reasons outlined by \citet{estebanetal04}. We have done a $\chi^2$ 
optimization of the values in the table, and we have obtained a $\chi^2$ parameter of 8.15, which
indicates a reasonable goodness of the fit for a system with nine degrees of freedom.

\subsection{Ionic Abundances from CELs}\label{cels}

Ionic abundances of N$^+$, O$^+$, O$^{++}$, Ne$^{++}$, S$^+$, S$^{++}$, Cl$^+$, Cl$^{++}$, 
Ar$^{++}$ and Ar$^{3+}$ have been determined from CELs, using the {\sc IRAF} package {\sc NEBULAR} 
\citep[except for Cl$^+$, see][]{garciarojasetal04}. 
Additionally, we have determined the ionic abundances of Fe$^{++}$ following the 
methods and data discussed in \citet{garciarojasetal04}. Ionic abundances are listed in 
Table ~\ref{celabun} and correspond to the mean value of the 
abundances derived from all the individual lines of each ion observed (weighted by their relative 
intensities).  

To derive the abundances for $t^2$ = 0.038 (see \S~\ref{tsq}) we used the abundances for $t^2$=0.00 and the 
formulation of \citet{peimbert67} and \citet{peimbertcostero69} for $t^2>$0.00. 
To derive abundances for other $t^2$ values it is possible to interpolate or extrapolate 
the values presented in Table~\ref{celabun}.

\setcounter{table}{5}
\begin{table}
\begin{minipage}{75mm}
\centering \caption{Ionic abundances from collisionally excited lines$^{\rm a}$.}
\label{celabun}
\begin{tabular}{lcc}
\noalign{\hrule} \noalign{\vskip3pt}
Ion & {\ts}=0.000 & {\ts}=0.038 $\pm$ 0.007 \\
\noalign{\vskip3pt} \noalign{\hrule} \noalign{\vskip3pt}
N$^{0}$&5.74 $\pm$ 0.06 & 5.88 $\pm$ 0.07 \\
N$^{+}$&7.25 $\pm$ 0.05 & 7.38 $\pm$ 0.06 \\
O$^{0}$&6.74 $\pm$ 0.06& 6.87 $\pm$ 0.06 \\
O$^{+}$&8.26 $\pm$ 0.07& 8.40 $\pm$ 0.08 \\
O$^{++}$& 7.81 $\pm$ 0.04 & 8.05 $\pm$ 0.06 \\
Ne$^{++}$& 6.81 $\pm$ 0.05 & 7.07 $\pm$ 0.07 \\
S$^{+}$&6.03 $\pm$ 0.05 & 6.15 $\pm$ 0.06 \\
S$^{++}$& 6.68 $\pm$ 0.07 & 6.95 $\pm$ 0.09 \\
Cl$^{+}$& 4.56 $\pm$ 0.05& 4.67 $\pm$ 0.05 \\
Cl$^{++}$& 4.85 $\pm$ 0.05& 5.08 $\pm$ 0.05 \\
Ar$^{++}$& 6.08 $\pm$ 0.04 & 6.28 $\pm$ 0.06 \\
Ar$^{3+}$& 3.42$^{+0.18}_{-0.30}$& 3.66$^{+0.18}_{-0.30}$ \\
Fe$^{++}$& 5.05 $\pm$ 0.06& 5.30 $\pm$ 0.08\\
\noalign{\vskip3pt} \noalign{\hrule} \noalign{\vskip3pt}
\end{tabular}
\begin{description}
\item[$^{\rm a}$] In units of 12+log(X$^m$/H$^+$).
\end{description}
\end{minipage}
\end{table}

Many {\ffeii} lines have been identified in our spectra, but all of them are severely affected 
by fluorescence effects \citep{rodriguez99b, verneretal00}. The only {\ffeii} line in the 
spectral range 3100 ${\rm \AA}$ to 10400 ${\rm \AA}$  
which is not affected by fluorescence effects is the {\ffeii} $\lambda8617$ ${\rm \AA}$ line, but 
unfortunately it is in one of our observational gaps.
We have measured {\ffeii} $\lambda7155$, a line which is not much affected by fluorescence effects 
\citep{verneretal00}, but it has a high observational error. Therefore, it was no possible to derive a reliable value 
of the Fe$^+$/H$^+$ ratio. The calculations for Fe$^{++}$ have been done with a 34 level 
model-atom that uses the collision strengths of \citet{zhang96} and the transition probabilities of 
\citet{quinet96}. We have used 6 {\ffeiii} lines that do not seem to be affected by line-blending and with errors less 
than 30 \%.
We find a mean value and a standard deviation of Fe$^{++}$/H$^+$=(1.115 $\pm$  
0.153)$\times$10$^{-7}$. Adding errors in {\te} and {\elecd} we finally obtain 
12+log(Fe$^{++}$/H$^+$)=5.05 $\pm$ 0.06. The value of the Fe$^{++}$ abundance for t$^2>$0.00 is also 
shown in Table ~\ref{celabun}.

\subsection{Ionic Abundances from Recombination Lines}\label{rls}

\setcounter{table}{6}
\begin{table}
\centering
\begin{minipage}{75mm}
\caption{C$^{++}$/H$^+$ abundance ratio from {\cii} lines}
\label{ciirl}
\begin{tabular}{ccccc}
\noalign{\hrule} \noalign{\vskip3pt}
\noalign{\hrule} \noalign{\vskip3pt}
& & $I$($\lambda$)/$I$(H$\beta$) & \multicolumn{2}{c}{C$^{++}$/H$^+$ ($\times$10$^{-5}$)} \\
Mult. & $\lambda_0$ & ($\times$10$^{-2}$) & A & B \\
\noalign{\vskip3pt} \noalign{\hrule} \noalign{\vskip3pt}
2& 6578.05$^{\rm a}$& 0.121 $\pm$ 0.015& 129 $\pm$ 15& 23 $\pm$ 3 \\
3& 7231.12& 0.038 $\pm$ 0.007& 1037 $\pm$ 197& 15 $\pm$ 3 \\
4& 3920.68& 0.046 $\pm$ 0.012& 400 $\pm$ 100& 130 $\pm$ 33 \\
6& 4267.26& 0.108 $\pm$ 0.013& 10 $\pm$ 1& {\bf 10 $\pm$ 1} \\ \hline 
& & & & \\ 
& Adopted& &\multicolumn{2}{c}{\bf 10  $\pm$  1 } \\ 
\noalign{\vskip3pt} \noalign{\hrule} \noalign{\vskip3pt}
\end{tabular}
\begin{description}
\item[$^{\rm a}$] Affected by a telluric emission line
\end{description}
\end{minipage}
\end{table}

\setcounter{table}{7}
\begin{table}
\centering
\begin{minipage}{90mm}
\caption{O$^{++}$/H$^+$ ratio from {\oii} permitted lines$^{\rm a}$}
\label{oiirl}
\begin{tabular}{cccccc}
\noalign{\hrule} \noalign{\vskip3pt}
\noalign{\hrule} \noalign{\vskip3pt}
& & $I$($\lambda$)/$I$(H$\beta$) & \multicolumn{3}{c}{O$^{++}$/H$^+$ ($\times$10$^{-5}$)$^{\rm b}$} \\
Mult. & $\lambda_0$ & ($\times$10$^{-2}$) & A & B & C \\
\noalign{\vskip3pt} \noalign{\hrule} \noalign{\vskip3pt}
1& 4638.85& 0.028 $\pm$ 0.008& 14 $\pm$ 4& 14 $\pm$ 4&– -- \\
& 4641.81& 0.028 $\pm$ 0.008& 13 $\pm$ 4& 12 $\pm$ 4&– -- \\ 
& 4649.14& 0.025 $\pm$ 0.008& 13 $\pm$ 4& 12 $\pm$ 4&– -- \\
& 4650.84& 0.025 $\pm$ 0.008& 11 $\pm$ 4& 11 $\pm$ 4&– -- \\
& 4661.64& 0.024 $\pm$ 0.008& 10 $\pm$ 4& 10 $\pm$ 4&– -- \\
& Sum& & 12 $\pm$ 1& {\bf 12 $\pm$ 1}&– -- \\ \hline
2& 4366.89& 0.092 $\pm$ 0.012& 25 $\pm$ 3& 18 $\pm$ 2& -- \\ \hline
19&4153.30& 0.024:& 794:/31:& 28:/28:& 28:/29: \\ \hline
& & & & & \\ 
& Adopted& &\multicolumn{3}{c}{\bf 12  $\pm$  1 } \\ 
\noalign{\vskip3pt} \noalign{\hrule} \noalign{\vskip3pt}
\end{tabular}
\begin{description}
\item[$^{\rm a}$] Only lines with intensity uncertainties lower than 40 \% have been considered
\item[$^{\rm b}$] Abundances of multiplet 1 have been corrected from NLTE effects (see text)
\end{description}
\end{minipage}
\end{table}

We have measured a large number of permitted lines of heavy element ions such as {\oi}, {\oii}, 
{\ci}, {\cii}, {\sii}, {\nitroi}, {\nii}, {\ari}, {\sili}, {\silii}, and {\fei}
most of them detected for the first time in S~311. Those permitted lines produced by recombination 
can give accurate determinations of ionic abundances because their intensities depend 
weakly on electron temperature and density. Unfortunately most of the permitted lines are 
affected by fluorescence 
effects or are blended with telluric emission lines making their intensities unreliable; 
also \citet{ruizetal03} have shown that, to
determine the abundances, it is important to measure all the lines of a
multiplet, because for low densities there could be an anomalous
distribition of the line intensities within the multiplet. 
Detailed discussions on the mechanism of formation of the permitted lines are in 
\citet[][and references therein]{estebanetal98, estebanetal04}.

\begin{figure}
\begin{center}
\epsfig{file=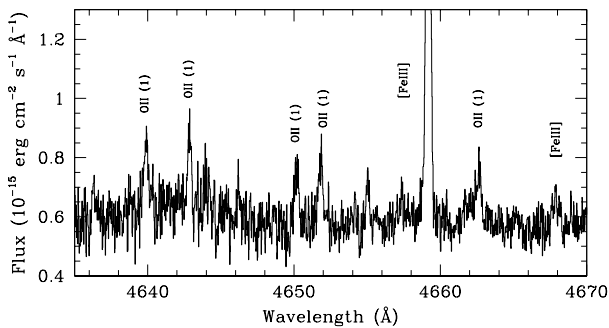, width=8. cm, clip=}
\caption{Section of the echelle spectrum showing the lines of multiplet 1 of {\oii}
(observed fluxes).
\label{oii}}
\end{center}
\end{figure}

We have been able of measuring ionic abundance ratios of O$^{++}$/H$^+$ and C$^{++}/$H$^+$ from pure 
recombination {\oii} and {\cii} lines respectively, from multiplet 1 for {\oii} (see Figure~\ref{oii}) 
and from multiplet 6 for {\cii} [see Figure 1 of \citet{estebanetal05}]. 
We have computed the abundances for {\te}(High)=9050 
K and {\elecd}=310 cm$^{-3}$. Atomic data and methodology 
for the C abundance are the same as in \citet{garciarojasetal04}. 
For the Multiplet
1 of O II it has been shown that for densities $n_e$ $<$ 10000 cm$^{-3}$ the upper
levels of the transition are not in LTE, and if one uses only one line to
determine the abundances one can have errors as large as a factor of 4 
\citep{ruizetal03}; instead we used the prescription presented by
\citet{apeimbertetal05} to calculate those populations; these
abundances show very good agreement among themselves and with the
abundance determined using the sum of all the lines, {\it ``Sum''}, that is not
expected to be affected by this effect. 
Tables \ref{ciirl} and \ref{oiirl} show the abundance ratios. 

\section{Temperature variations}\label{tsq}

It is well known that under the assumption of a constant temperature, RLs of heavy 
element ions yield 
higher abundance values relative to hydrogen than CELs in {\hii} regions \citep[e.g.][and references
therein]{peimbertetal93, estebanetal98, esteban02, estebanetal04, garciarojasetal04, tsamisetal03}. 
\citet{torrespeimbertetal80} proposed the 
presence of spatial temperature fluctuations (parametrized by {\ts}) as the cause of this 
discrepancy, because CELs and RLs emissivities have different dependences on the electron 
temperature. On the other hand, \citet{peimbert71} proposed that there is a dichotomy between {\te} derived from 
the {\foiii} lines and from the hydrogen recombination continuum discontinuities, which is strongly 
correlated with the discrepancy between CEL and RL abundances, so the comparison between electron
temperatures obtained from both methods is an additional
indicator of {\ts}. A complete formulation of temperature fluctuations has been developed by 
\citet{peimbert67}, \citet{peimbertcostero69} 
and \citet{peimbert71}. We have assumed a two-zone ionization scheme, and we have followed 
the re-formulation of \citet{peimbertetal00} and \citet{apeimbertetal02} to derive the value of 
{\ts} comparing {\te}($Bac$)
and {\te}($Pac$) with the combination of {\te}({\foii}) and {\te}({\foiii}), {\ts}($Bac-FL$) and
{\ts}($Pac-FL$), respectively, using equation (A1) of
\citet{apeimbertetal02}. Table~\ref{t2} shows the different {\ts} values obtained as well as the adopted value,
{\ts}=0.038 $\pm$ 0.007 which is the weighted average of O$^{\rm ++}$ and He$^{\rm +}$ values which are rather
consistent and show the lowest uncertainties.

\setcounter{table}{8}
\begin{table}
\begin{minipage}{75mm}
\centering \caption{{\ts} parameter}
\label{t2}
\begin{tabular}{cc}
\noalign{\hrule} \noalign{\vskip3pt}
Method & {\ts} \\
\noalign{\vskip3pt} \noalign{\hrule} \noalign{\vskip3pt}
O$^{\rm ++}$ (R/C)& 0.040 $\pm$ 0.008 \\
He$^{\rm +}$ & 0.034 $\pm$ 0.010 \\
Bac--FL & 0.002 $\pm$ 0.022\\ 
Pac--FL & 0.019 $\pm$ 0.026\\ \hline
Adopted & 0.038 $\pm$ 0.007\\ 
\noalign{\vskip3pt} \noalign{\hrule} \noalign{\vskip3pt}
\end{tabular}
\end{minipage}
\end{table}

\setcounter{table}{9}
\begin{table}
\begin{minipage}{75mm}
\centering \caption{Continuum determinations$^{\rm a}$}
\label{cont}
\begin{tabular}{cccc}
\noalign{\hrule} \noalign{\vskip3pt}
$\lambda$(\AA ) &\multicolumn{3}{c}{log ($j$($\lambda$))/$I$({\hb})} \\
\noalign{\vskip3pt} \noalign{\hrule} \noalign{\vskip3pt}
& Atomic & Observed & Scattered light \\
\noalign{\vskip3pt} \noalign{\hrule} \noalign{\vskip3pt}
3640 & --2.235 &  --2.106 $\pm$ 0.009 &  --2.699 $\pm$ 0.033 \\
3670 & --3.057 &  --2.469 $\pm$ 0.025 &  --2.598 $\pm$ 0.033 \\
4110 & --3.185 &  --2.577 $\pm$ 0.024 &  --2.700 $\pm$ 0.031 \\
4350 & --3.213 &  --2.676 $\pm$ 0.022 &  --2.825 $\pm$ 0.031 \\
4850 & --3.245 &  --2.767 $\pm$ 0.015 &  --2.943 $\pm$ 0.022 \\
6650 & --3.310 &  --2.870 $\pm$ 0.013 &  --3.066 $\pm$ 0.020 \\
8190 & --3.354 &  --3.059 $\pm$ 0.004 &  --3.367 $\pm$ 0.009 \\
8260 & --3.883 &  --3.262 $\pm$ 0.003 &  --3.381 $\pm$ 0.004 \\
\noalign{\vskip3pt} \noalign{\hrule} \noalign{\vskip3pt}
\end{tabular}
\begin{description}
\item[$^{\rm a}$] in units of (\AA$^{-1}$)
\end{description}
\end{minipage}
\end{table}

On the other hand, as it can be seen from Table~\ref{tsq}, {\ts}($Bac-FL$) and {\ts}($Pac-FL$) 
are significantly lower than 
the other values. One possible explanation is that the nebular continuum could be affected by dust 
scattered light. To explore that possibility we have derived the atomic continua contribution, which includes the free-free 
and free-bound 
continua of the hydrogen and helium atoms and the two-quantum continuum, from the computations of
\citet{brownmathews70} for {\te}=7620 K, {\elecd}=310 cm$^{-3}$ and He$^+$/H$^+$=0.0795 (see
section~\ref{heabund}). The temperature of 7620 K we have assumed is that which implies {\ts}($Bac-FL$) and
{\ts}($Pac-FL$) equal to the final adopted value. Table~\ref{cont} shows the observed and the 
expected atomic continua, and the derived scattered
light contribution. From these data it is easy to derive the contribution of dust scattered light to
the continuum near the Balmer lines and the Balmer and Paschen jumps. 
Including observational 
uncertainties, we have derived that a contribution  between 10\% and 30\% to the Balmer 
jump of the scattered continuum integrated light is enough to explain the high {\te}($Bac$)
obtained from the discontinuity and, therefore, the lower {\ts}($Bac-FL$). The shape of the 
spectral distribution of the scattered light is then similar to a B3 V star.
Assuming that the ionising star of S~311 --HD~64315-- is a main sequence O6e, we have modelled
the optical flux distribution using FASTWIND, an spherically symmetric, 
NLTE model atmosphere code \citep{santolayareyetal97, pulsetal05} and the corresponding 
$T_{\rm eff}$ and $log g$ derived from \citet{martinsetal05} calibrations. 
From this model, we have 
obtained that the stellar Balmer jump is about 10\% of the stellar continuum. 
Anyway, our slit position is far from HD~64315 (see Figure~\ref{halpha}) and  
other late O and early B stars may contribute to the continuum scattered light, so the  
contribution of the discontinuities would be even higher, and therefore, the  
nebular temperature determination would decrease. In fact, \citet{ feinsteinvazquez89} 
list a B8.5 star near our slit position (their star labelled as NGC 2467--12), whose 
position is also indicated in Figure 1.  
It can be seen that, as expected, the continuum scattered light increases monotonically
towards the blue. 
A more detailed study of the properties of the dust in S311 (albedo, 
reddening and geometrical distribution) 
is needed to solve this problem in an appropiate way,
but this is outside the scope of this paper.

\begin{figure}
\begin{center}
\epsfig{file=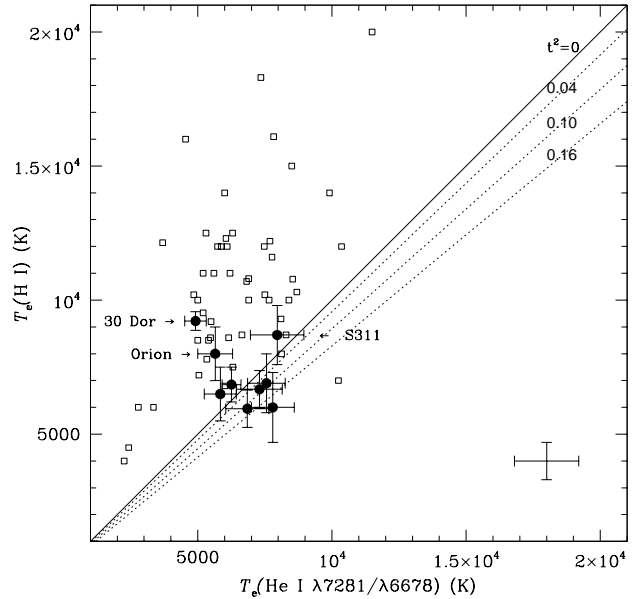, width=9. cm, clip=}
\caption{$T_e$({\hei}) versus $T_e$({\hi}). This figure is like Figure 5 of \citet{zhangetal05} for 
PNe (open squares),
including data for {\hii} regions (filled circles). The solid diagonal line is a y = x plot, and the dotted
lines show the variations of T$_e$({\hi}) as a function of T$_e$({\hei}) for different {\ts}
values. The cross on the lower right part of the diagram 
represent typical uncertainties for PNe data.
\label{zhang}}
\end{center}
\end{figure}

Recently, \citet{zhangetal05} have computed {\te}({\hei}) for 48 planetary nebulae 
from {\hei} recombination line ratios ({\hei $\lambda$7281/$\lambda$6678}). These authors have found 
that temperature fluctuations do not predict the general behaviour of the {\te}({\hei}) vs. {\te}({\hi}) 
diagram. Using the expression given by \citet{zhangetal05} we have derived {\te}({\hei})=7960 $\pm$ 1000 K, 
for S~311, which is lower, but consistent within the errors with our {\te}({\hei}) and {\te}({\hi}). 
In Figure~\ref{zhang} we have compared results from \citet{zhangetal05} for 
planetary nebulae (PNe) with our results for {\hii} regions, in which we have included S311 (this paper), 
NGC 3576 \citep{garciarojasetal04} and still unpublished echelle VLT spectra of M8, M17, 
M20, M16 and NGC 3603. It can be seen that most of the 
{\hii} regions, except for the Orion nebula \citep{estebanetal04} and 30 Dor \citep{apeimbert03}, do not follow 
the behaviour of the bulk of PNe and do not contradict the 
temperature fluctuations paradigm.
\citet{zhangetal05} solve the problem of the anomalous position of PNe in the 
$T_e$({\hei}) vs. $T_e$({\hi}) diagram proposing the presence of a small amount of H-
deficient material for most of the sample nebulae, in the context of the scenario of H-
deficient inclusions constructed by \citet{pequignotetal03}. That scenario seems to explain
successfully the large abundance discrepancies reported in some PNe. 
However, Figure~\ref{zhang} suggests that the behaviour of {\hii} regions is qualitatively different 
to that of PNe and that the position of most {\hii} regions in the diagram is consistent 
with the classical {\ts} paradigm. This result is in agreement with the 
suggestions outlined by \citet{esteban02} who propose that the processes producing  
abundance discrepancies in {\hii} regions and PNe -at least the extreme cases of PNe- 
could be different. In the case of {\hii} regions the observational evidences are still consistent 
with the {\ts} scheme, while in the case of the extreme PNe this scheme seems to fail.

\section{Total abundances}\label{totabun}

\setcounter{table}{10}
\begin{table*}
\begin{minipage}{125mm}
\centering \caption{Total Gaseous Abundances.}
\label{total}
\begin{tabular}{lccccc}
\noalign{\hrule} \noalign{\vskip3pt}
 & \multicolumn{2}{c}{S~311 (this work)}  & Orion$^{\rm a}$ & \\
Element& {\ts}=0.000&{\ts}=0.038 $\pm$ 0.007 & {\ts}=0.022 $\pm$ 0.002& Sun$^{\rm b}$ & S~311$-$Orion \\
\noalign{\vskip3pt} \noalign{\hrule} \noalign{\vskip3pt}
He&10.99 $\pm$ 0.02 & 10.97 $\pm$ 0.02& 10.988 $\pm$ 0.003& 10.98 $\pm$ 0.02 & --0.018\\
C&8.38 $\pm$ 0.05 & 8.38 $\pm$ 0.07& 8.42 $\pm$ 0.02& 8.39 $\pm$ 0.05& --0.04 \\
N&7.43 $\pm$ 0.06 & 7.61 $\pm$ 0.07& 7.73 $\pm$ 0.09& 7.78 $\pm$ 0.06& --0.12 \\
O&8.39 $\pm$ 0.05& 8.56 $\pm$ 0.06& 8.67 $\pm$ 0.04& 8.66 $\pm$ 0.05& --0.11 \\
O$^{\rm c}$&8.54 $\pm$ 0.10& 8.57 $\pm$ 0.05 & 8.65 $\pm$ 0.03& " & --0.08\\
Ne& 7.79 $\pm$ 0.13 & 7.98 $\pm$ 0.14 & 8.05 $\pm$ 0.07& 7.84 $\pm$ 0.06& --0.07 \\
S&6.77 $\pm$ 0.06 & 7.02 $\pm$ 0.08& 7.22 $\pm$ 0.04& 7.14 $\pm$ 0.05& --0.20 \\
Cl& 5.03 $\pm$ 0.06& 5.22 $\pm$ 0.07 & 5.28 $\pm$ 0.04& 5.23 $\pm$ 0.06& --0.06 \\
Ar& 6.43 $\pm$ 0.07 & 6.56 $\pm$ 0.08 & 6.62 $\pm$ 0.05& 6.18 $\pm$ 0.08& --0.06 \\
Fe& 5.17 $\pm$ 0.11& 5.44 $\pm$ 0.13& 6.23 $\pm$ 0.08& 7.45 $\pm$ 0.05& --0.79 \\
\noalign{\vskip3pt} \noalign{\hrule} \noalign{\vskip3pt}
\end{tabular}
\begin{description}
\item[$^{\rm a}$] Gas abundances from \citet{estebanetal04}
\item[$^{\rm b}$] \citet{christensen98, asplundetal05}
\item[$^{\rm c}$] O$^{++}$/H$^+$ from RLs and O$^+$/H$^+$ from CELs and t$^2$
\end{description}
\end{minipage}
\end{table*}

We have adopted a set of ionization correction factors (ICFs) to 
correct for the unseen ionization
stages and then derive the total gaseous abundances of the 
elements we have studied. We have 
adopted the ICF scheme used by \citet{garciarojasetal04} for 
all the elements except for carbon, neon, chlorine 
and iron. 

The absence of {\heii} lines in our spectra indicates that He$^{++}$/H$^+$ is 
negligible. However, the total helium abundance has to be corrected for the presence 
of neutral helium. Based on the ICF(He$^0$) given by \citet{peimbertetal92}, and our data 
ICF(He$^0$) amounts to 1.22 $\pm$ 0.05 for {\ts} = 0.00 and 1.16 $\pm$ 0.04 
for {\ts} $>$ 0.00.

We have derived the O/H ratio both from CELs 
and from the combination of
O$^{++}$/H$^+$ ratio from RLs and O$^{+}$/H$^+$ ratio 
from CELs and the assumed {\ts}. 

For carbon we have adopted the ICF derived from photoionization models by 
\citet{garnettetal99}. 

For neon the ICF proposed by \citet{peimbertcostero69} given by:

\begin{equation}
\frac{N(\rm Ne)}{N(\rm H)} =
            \left( \frac{N({\rm O^+})+N(\rm O^{++})}{N(\rm O^{++})} \right)
            \frac{N(\rm Ne^{++})}{N(\rm H^+)}.
\end{equation}

has been generally used. Nevertheless this ICF underestimates the Ne/H
abundance for nebulae of low degree of ionization because a considerable
fraction of Ne$^+$ coexists with O$^{++}$ 
\citep[see][]{torrespeimbertpeimbert77,peimbertetal92}. For
S~311 based on the O$^+$/O ratio and  the data by \citet{torrespeimbertpeimbert77}, 
we estimate that the ICF(Ne) should be 0.4  $\pm$  0.1 dex
higher than that provided by the previous equation.

We have measured lines of two ionization stages of chlorine: 
Cl$^{+}$ and Cl$^{++}$. The Cl abundance has been assumed to be equal to 
the sum of these ionic abundances without taking into account Cl$^{3+}$ 
fraction. This assumption seems reasonable taking into account the small 
Cl$^{3+}$/Cl$^{++}$ ratio found for M17 \citep[$\sim$0.03, see][]{estebanetal99a}, 
for the Orion nebula \citep[$\sim$0.04, see][]{estebanetal04}, and for NGC3576 
\citep[$\sim$0.02, see][]{garciarojasetal04}, and the lower ionization degree of S~311 
with respect to those nebulae.

We have measured lines of two stages of ionization of iron: Fe$^+$ and Fe$^{++}$, but in 
\S~\ref{cels} we have shown that Fe$^+$/H$^+$ ratio is not reliable. Recently, 
\citet{rodriguezrubin05} have derived an ICF from a least-squares fit 
to the results of a set of models in which it is represented 
$\chi$(O$^+$)/$\chi$(Fe$^{++}$), the ratio of ionization fractions of O$^+$ and Fe$^{++}$, 
respectively, as a function of the degree of ionization given by O$^+$/O$^{++}$. 
The ionization correction scheme they have derived is as follows:

\begin{equation}
\frac{N({\rm Fe})}{N({\rm H})} = 0.9\left[\frac{N({\rm O}^{+})}{N({\rm O}^{++})}\right]^{0.08}\times\frac{N({\rm Fe})^{++}}{N({\rm O})^{+}}\times\frac{N({\rm O})}{N({\rm H})}.
\end{equation}

In Table~\ref{total} we show the total abundances obtained for our slit 
position in S~311 for {\ts}=0.00 and {\ts}=0.038 $\pm$ 0.007. 

\section{Deuterium Balmer lines}\label{deut}

\begin{figure}
\begin{center}
\epsfig{file=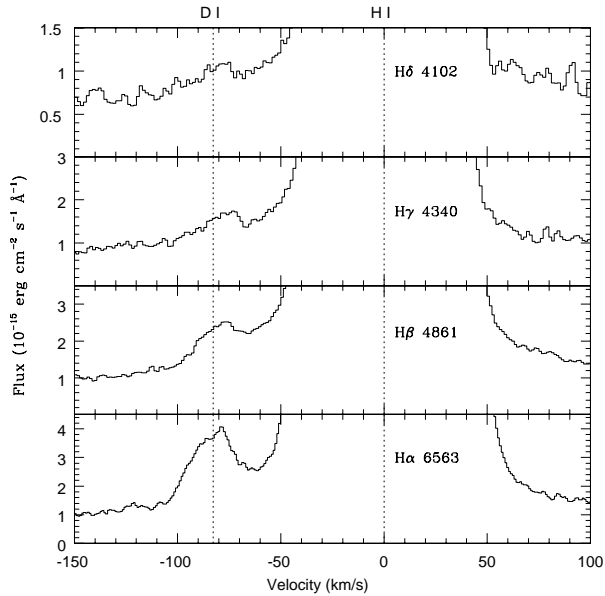, width=9. cm, clip=}
\caption{Wings of {\ha} to H$\delta$ in S311. The lines are centred at 0 km s$^{-1}$ 
velocity. The dotted line of the left correspond to the average wavelength adopted 
for the {\di} lines.
\label{deuterium}}
\end{center}
\end{figure}

We have detected, for the first time in an {\hii} region outside the solar circle, 
the four brightest deuterium Balmer lines, D$\alpha$, D$\beta$, D$\gamma$ and D$\delta$ 
as very weak lines in the blue wings of the corresponding {\hi} Balmer lines (see 
Figure~\ref{deuterium}). 
The apparent shift in radial velocity of these weak lines with respect to the hydrogen ones is $-$82.7 km 
s$^{-1}$ (see Table~\ref{deuchar}), which is in excellent agreement with the isotopic shift of deuterium, 
$-$81.6 km s$^{-1}$. We have discarded these weak features as high velocity components of 
hydrogen for the following reasons:

\begin{itemize}

\item We have not found blue-shifted features in the wings of the brightest {\fnii}, 
{\foii} and {\foiii} lines, indicating that the faint features in the blue wings of 
{\hi} lines can not come from emission of high-velocity ionized material. 

\item They are narrower (FWHM $\leq 10$ km s$^{-1}$) than hydrogen lines 
(FWHM $\sim 20$ km s$^{-1}$). FWHM has been derived from gaussian fits, after correcting
from the underlying blue wing of the corresponding Balmer line, and
after quadratic subtraction of the instrumental point-spread function. Although the relatively 
low velocity resolution of our spectra, which is not enough to derive precisely
the value of the thermal width of the deuterium Balmer lines (colons in the FWHM 
indicates high uncertainties, and the low values of the FWHM of the deuterium lines are 
because deuterium line widths are of the order of the instrumental one), it is sufficient to 
compare qualitatively with the value of the width of hydrogen Balmer lines 
(see Table~\ref{deuchar}). This result supports the idea that deuterium lines
arise from a cold material with smaller thermal velocity, probably the 
photon dominated region or PDR \citep{hebrardetal00a}.

\end{itemize}
 
The detection and identification of the deuterium Balmer lines in an {\hii} region
were first reported by \citet{hebrardetal00a} from VLT/UVES data; they confirmed the detection 
and identification of deuterium Balmer lines (up to D$\eta$) in the Orion nebula.
Subsequently, \citet{hebrardetal00b} published the detection and identification of 
at least D$\alpha$ and D$\beta$ in four additional {\hii} regions (M8, M16, M20 and 
DEM S~103 in SMC), and confirmed fluorescence as the main excitation mechanism 
of {\di}, recombination being negligible.
The {\di}/{\hi} ratios presented in Table~\ref{deuchar} correspond to the intensity ratios 
and are upper
limits to the abundance ratio because in addition to recombination the {\di} line
intensities include the fluorescence contribution that is considerably larger 
than the recombination one.

\setcounter{table}{11}
\begin{table}
\begin{minipage}{85mm}
\centering \caption{Deuterium Balmer line characteristics in S~311.}
\label{deuchar}
\begin{tabular}{ccccc}
\noalign{\hrule} \noalign{\vskip3pt}
Line & {\di} Isotopic & FWHM {\di} & FWHM {\hi} &{\di}/{\hi} ratio \\
     &  shift (km s$^{-1}$) & (km s$^{-1}$)   & (km s$^{-1}$) & ($\times$10$^{-4}$)  \\
\noalign{\vskip3pt} \noalign{\hrule} \noalign{\vskip3pt}
$\alpha$ & $-83.6$ & 8:   & 20 &7.0 $\pm$ 0.7 \\
$\beta$  & $-81.4$ & $<$4:     & 20 & 8.5 $\pm$ 1.6 \\
$\gamma$ & $-80.1$ & $\sim$0:  & 20 & 9.4 $\pm$ 2.1 \\
$\delta$ & $-85.5$ & 17:   & 21 & 12.9 $\pm$ 3.9 \\
\noalign{\vskip3pt} \noalign{\hrule} \noalign{\vskip3pt}
\end{tabular}
\end{minipage}
\end{table}

On the other hand, \citet{odelletal01} presented observations and a model for the emission of 
the deuterium lines in Orion, and concluded that they are produced by fluorescent excitation 
of the upper energy states by the far-UV radiation of the ionising star. In S 311 the 
D$\alpha$/{\ha} and D$\beta$/{\hb} ratios are somewhat larger than in 
the case of the Orion nebula. In the light of the model outlined by \citet{odelletal01}, 
and taking into account that the spectral types of the ionizing stars of both nebulae 
are similar, this can be due to an additional contribution of UV radiation from 
other nearby cooler stars or to a lower UV grain extinction in S~311. On the other hand, the 
comparison of the Balmer decrements of the hydrogen and deuterium lines  
observed in our spectra follow closely the standard fluorescence model by 
\citet[][see their Figure 13]{odelletal01} for the Orion nebula. 

\section{Discussion}\label{discus}

Few optical spectrophotometric studies of S~311 have been published in the literature.
The most complete are those presented by \citet{hawley78} and \citet{peimbertetal78}. 
Unfortunately, these papers do not study the same slit position as ourselves (see \S~\ref{lin}), and 
the ionic abundances they have obtained are also very different. In fact, the ratio O$^+$/O$^{++}$, 
which is an indication of the ionization degree of the nebula is very similar in these 
works ranging from 
0.89 to 1.00; on the contrary, our O$^+$/O$^{++}$=2.8 (for {\ts}=0.00) is much larger, 
pointing out that 
our slit position is nearer the ionization front of the nebula, where the O$^+$/O$^{++}$ ratio 
increases significantly. In fact, as it 
can be seen in Figure~\ref{halpha}, our slit position is on the brightest part 
of the nebula, coinciding with a filament or an ionization front.

On the other hand, we can make comparisons with spectrophotometric data of \citet{shaveretal83}. 
Their slit position 2 is closer to ours. 
These authors only derived ionic abundances for 5 species: He$^{+}$, O$^{++}$, N$^{+}$, 
S$^{+}$ and S$^{++}$. They assumed temperatures much lower than ours, about 800 K lower for 
the high ionization species, and more than 1500 K lower for low ionization ones. 
This implies, in general, an overestimation 
of the abundances with respect to us, except for O$^{++}$, which would be underestimated. 
We have derived the abundances from \citet{shaveretal83} line intensities 
using the atomic data listed in Table~\ref{atom} 
and we have obtained a good agreement 
between them and our results, except for S$^{++}$/H$^{+}$ which is 0.21 dex lower than our derived abundance.  
Although \citet{shaveretal83} did not quote errors in the line intensities, this fact is possibly due to 
uncertainties in the measured flux of the {\fsiii}$\lambda$6312 line. 
Taking into account the faintness of this line in the spectrum of S~311 (1\% 
of H$\beta$), we have estimated their error in the flux measurement of {\fsiii}$\lambda$6312 
in about 50\%, which is enough to explain the 
difference between the derived abundances. For the He$^{+}$/H$^{+}$ ratio the highest temperature assumed and to 
consider temperature fluctuations make us to derive a value 0.13 dex higher than the one derived by \citet{shaveretal83}.

Obtaining deep good-quality spectra of {\hii} regions located outside the solar circle 
is of great importance for deriving radial abundance gradients in the Galactic 
disk. {\hii} regions in this part of the Galaxy are scarce and usually very faint 
\citep[][]{russeil03}. In fact, part of the abundance data presented in this paper 
(C, N and O abundances) have been included in recent works devoted to the 
calculation and modelling of abundance gradients \citep{estebanetal05,carigietal05b}.

In Table~\ref{total} we compare S~311 and Orion nebula gas abundances and the solar 
values. For the Sun: He comes from \citet{christensen98} and the  
rest of elements from \citet{asplundetal05}.  
As expected, the total abundances of S~311 are somewhat lower than those of the Orion nebula 
and the Sun because of the existence of Galactic radial abundance gradients and the larger 
Galactocentric distance of S~311. 
Interestingly, the Fe abundance is very different in both nebulae, this could be due to  
their different dust-depletion factors. 

It is important to appreciate that nebulae are 3-D objects and the spectrum 
really corresponds to the integral of the emission contained in the column of gas 
covered by the slit area. This implies that the emission comes from a range of 
densities, degrees of ionization, temperatures and even extinctions within the column. 
However, our observations are limited to a small area covering  
a bright rim, probably coincident with a filament or an ionization front, precisely the 
brightest part of the nebula, where we expect the 3-D effects should be minimum. Nevertheless, 
it would be interesting to make realistic 3-D \citep{ercolanoetal03} or pseudo 3-D models 
\citep{morissetetal05} for Galactic {\hii} regions, combined with medium-high resolution 
long-slit spectroscopic data and narrow-band imaging in different emission line filters.
On the other hand, 3-D effects should be much more severe in the case of extragalactic HII 
regions, where a small slit area covers a enormous volume 
of gas (several orders of magnitude larger than in Galactic {\hii} regions). 
In a very recent paper, \citet{pilyugin05} has found that our data for S~311 do not fit 
his strong-line diagnostic diagrams for the empirical derivation of chemical abundances. 
This deviation is clearly due to the fact that the line fluxes accepted by the slit are not 
representative of the nebula as a whole. This effect has to be taken into account when 
spatially resolved observations of small zones of a nebula are used to apply empirical methods for the 
derivation of abundances.

\section{Conclusions}\label{conclu}

We present echelle spectroscopy in the 3100--10400 ${\rm \AA}$ range of 
the brightest zone of the Galactic {\hii} region S~311. We have measured the intensity of 263 emission lines.
This is the deepest spectra ever taken for a 
Galactic {\hii} region located outside the solar circle. 

We have derived the physical conditions of S~311 making use of several line intensities and continuum ratios. 
The chemical abundances have been derived using CELs for a large number of ions. 
We have determined also, for the first time in S~311, the C$^{++}$ and O$^{++}$ abundances from RLs. 
We have obtained an average {\ts}=0.038 $\pm$ 0.007 both by comparing the O$^{++}$ ionic abundance 
derived from RLs to those derived from CELs and by applying a chi-squared method which minimizes the 
dispersion of He$^{+}$/H$^{+}$ ratio from individual lines. The adopted average value has been used to 
derive the abundances determined from CELs.

We have compared {\te}({\hei}) {\it vs.} {\te}({\hi}) for S~311 and other Galactic 
{\hii} regions finding that the {\ts} paradigm is consistent with the observed 
behaviour of most of the objects, in contrast with what has been found for PNe. 
This result suggests that the process or processes that produce the abundance 
discrepancies in both kinds of objects could be different.

We have detected four deuterium Balmer emission lines in S~311. These are the first detections of 
these lines in this object. Comparison with previous observations and with models lead us to 
support that fluorescence seems to be the most probable excitation mechanism 
of these lines.

\section*{Acknowledgments}

This work is based on observations collected at the European Southern Observatory, Chile, 
proposal number ESO 68.C-0149(A). We thank the annonymous referee for useful comments. 
We would like to thank A. R. L\'opez-S\'anchez 
for providing us the H$\alpha$ image of S~311. JGR would like to thank S. Sim\'on-D\' iaz for providing us
the results of stellar modelling for HD~64315, and 
E. P\'erez-Montero for fruitful discussions and excellent suggestions. JGR and CE would like to thank 
the members of the Instituto de Astronom\'ia, UNAM, for their
always warm hospitality. This work has been partially funded by the Spanish 
Ministerio de Ciencia y Tecnolog\'{\i}a (MCyT) under projects AYA2001-0436 and 
AYA2004-07466. MP received partial support from DGAPA UNAM (grant IN 114601). 
MTR received partial support from FONDAP(15010003) and
Fondecyt(1010404). MR acknowledges support from Mexican CONACYT project J37680-E.


\end{document}